\documentstyle[12pt]{article}
\def\theequation{\arabic{section}.\arabic{equation}}
\begin{document}
\newcommand{\ti}{\theta^{1,0\;\und i}}
\newcommand{\tl}{\theta^{1,0\;\und l}}
\newcommand{\tk}{\theta^{1,0\;\und k}}
\newcommand{\tx}{\theta^{1,0}_{\und i}}
\newcommand{\ta}{\theta^{0,1\;\und a}}
\newcommand{\tb}{\theta^{0,1\;\und b}}
\newcommand{\tc}{\theta^{0,1\;\und c}}
\newcommand{\ty}{\theta^{0,1}_{\und a}}
\newcommand{\da}{D^{2,0}}
\newcommand{\db}{D^{0,2}}
\newcommand{\du}{D^{0,0}_u}
\newcommand{\dv}{D^{0,0}_v}
\newcommand{\qa}{q^{1,0\;\und a}}
\newcommand{\qb}{q^{0,1\;\und i}}
\newcommand{\qc}{q^{\und i\;\und a}}
\newcommand{\nn}{\nonumber}
\newcommand{\be}{\begin{equation}}
\newcommand{\bea}{\begin{eqnarray}}
\newcommand{\eea}{\end{eqnarray}}
\newcommand{\ee}{\end{equation}}
\newcommand{\eps}{\varepsilon}
\newcommand{\und}{\underline}
\newcommand{\p}[1]{(\ref{#1})}

\begin{titlepage}
\begin{flushright}
JINR-E2-2004-28 \\
hep-th/0403130
\end{flushright}
\vskip 1.0truecm
\begin{center}
{\large \bf DIVERSITY OF OFF--SHELL TWISTED (4,4) MULTIPLETS \\

\vspace{0.2cm}
IN SU(2) x SU(2) HARMONIC SUPERSPACE}
\vglue 1.5cm
{\bf E. Ivanov and A. Sutulin}
\vglue 1.5cm
{\it Bogoliubov Laboratory of Theoretical Physics,
JINR,\\
141 980 Dubna, Moscow Region, Russia}\\

\vspace{0.2cm}
{\tt eivanov,sutulin@thsun1.jinr.ru}
\end{center}

\vspace{2cm}

\begin{abstract}
We elaborate on four different types of twisted ${\cal N}=(4,4)$
supermultiplets in the $SU(2) \times SU(2)$, $2D$ harmonic superspace.
In the conventional ${\cal N}=(4,4)$, $2D$ superspace they are described by
the superfields $\hat q^{\,i\,a}$\,, $\hat q^{\,\und i\, a}$\,, $\hat
q^{\,i\, \und a}$\,, $\hat q^{\,\und i\, \und a}$ subjected to proper
differential constraints, $(i, \und i, a, \und a)$ being
the doublet indices of four groups $SU(2)$ which form the full
R-symmetry group $SO(4)_L\times SO(4)_R$ of ${\cal N}=(4,4)$ supersymmetry.
We construct the torsionful off--shell sigma model actions for each type
of these multiplets, as well as the corresponding invariant mass terms,
in an analytic subspace of the $SU(2) \times SU(2)$ harmonic superspace.
As an instructive example, ${\cal N}=(4,4)$ superconformal extension of the
$SU(2) \times U(1)$ WZNW sigma model action and its massive deformation
are presented for the multiplet $\hat q^{\,i\, \und a}$\,.
We prove that ${\cal N}=(4,4)$ supersymmetry requires the general sigma model action
of pair of different multiplets to split into a sum of sigma model actions
of each multiplet. This phenomenon also persists if a larger number of
non-equivalent multiplets are simultaneously included. We show that different
multiplets may interact with each other only through mixed mass terms
which can be set up for multiplets belonging to ``self-dual'' pairs
$(\hat q^{\,i\,a}\,, \hat q^{\,\und i\, \und a})$ and
$(\hat q^{\,\und i\, a}\,, \hat q^{\,i\, \und a})$\,.
The multiplets from different pairs cannot interact at all.
For a ``self-dual'' pair of the twisted multiplets we give the most general
form of the on-shell scalar potential.
\end{abstract}
\end{titlepage}

\section{Introduction}
The interest in ${\cal N}=(4,4)$ supersymmetric two-dimensional sigma models with torsion
has a long history and is mainly motivated by the important role the corresponding
target spaces play in string theory and AdS/CFT correspondence
(see e.g. \cite{1}-\cite{5} and refs. therein).
The first example of such a model, ${\cal N}=(4,4)$ supersymmetric (and superconformal)
extension of $SU(2)\times U(1)$ WZNW model, was discovered in ref. \cite{IK} as a special
case of ${\cal N}=(4,4)$ super-Liouville system (see also \cite{IKL}).
A more general class of these sigma models was presented
in \cite{GHR,HP}. In \cite{belg1}, ${\cal N}=(4,4)$ superextensions of other group-manifold
WZNW models were constructed, and the exhaustive list of group manifolds for which
such extensions exist was given (they are those admitting a quaternionic structure).
Superfield formulations of ${\cal N}=(4,4)$ models were given in ${\cal N}=(2,2)$ superspace
\cite{GHR}, in ordinary ${\cal N}=(4,4)$ superspace \cite{GI,G1}, in the projective
superspace \cite{BLR,RSS} and in the ${\cal N}=(4,4)$, $SU(2)\times SU(2)$ bi-harmonic
superspace \cite{IS} - \cite{bi1}. The ${\cal N}=(4,4)$ superfield formulations
are most appropriate, as they make manifest and off-shell the full amount of
the underlying supersymmetry. As argued in \cite{IS}, the bi-harmonic formulations
are especially advantageous because they manifest not only supersymmtery but also
internal R-symmetry $SU(2)_L\times SU(4)_R$ of ${\cal N}=(4,4)$, $2D$ Poincar\'e  supersymmetry,
and allow one to control how this symmetry is broken in various sigma model actions.

The bosonic target geometry of general ${\cal N}=(4,4)$ supersymmetric sigma models of
the considered type was studied e.g. in \cite{GHR,belg,hp}.
One of its versions is characterized by two sets of mutually commuting
covariantly constant quaternionic structures.
Any sigma model of this sort can be  described off shell by the ``twisted''
${\cal N}=(4,4)$, $2D$ supermultiplets with the off-shell content $({\bf 4, 8, 4})$. As  for
${\cal N}=(4,4)$ sigma models with {\it non-commuting} left and right complex structures,
which involve e.g. most of the group manifold ${\cal N}=(4,4)$ WZNW sigma models
(with exception of $SU(2)\times U(1)$ and some other product manifolds including
$SU(2)$ and $U(1)$ as the factors \cite{belg1,belg}),
not too much is known about their superfield description. What is certainly known is
that it is impossible to formulate them in terms of ${\cal N}=(4,4)$ twisted multiplets
alone \cite{semi}. In terms of some other multiplets, models of this kind were discussed in the framework
of ${\cal N}=(2,2)$ superspace \cite{BLR,ds,semi} and in the bi-harmonic ${\cal N}=(4,4)$
superspace \cite{EI} (in the latter case, superfields with infinitely many auxiliary fields have
to be involved, and the Poisson structures on the target space naturally appear).

Yet in the case of ${\cal N}=(4,4)$ sigma models based on twisted multiplets there
is a subtlety related to the existence of few types of these multiplets which
differ in the transformation properties of their component fields with respect
to the full R-symmetry group $SO(4)_L\times SO(4)_R$ of ${\cal N}=(4,4)$, $2D$
Poincar\'e superalgebra. This degeneracy of twisted multiplets was first
noticed in \cite{GI,GKet,ivan1}.
\footnote{As observed in \cite{GKet}, even further proliferation
of non-equivalent twisted multiplets can be achieved by grading their
components in different ways under $2D$ space-time parity.}
It is clearly seen just in the ${\cal N}=(4,4)$ superfield language where the various
twisted multiplets are represented by the properly constrained superfields \cite{ivan1}
\be
\hat q^{\,i\,a}\,, \quad \hat q^{\,i\, \und a}\,, \quad \hat q^{\,\und i\, a}\,, \quad
\hat q^{\,\und i\, \und a}\,. \label{base}
\ee
Here,  the external doublet indices $i, \und i$ and $a, \und a$
refer to two left and two right $SU(2)$ constituents of
the R-symmetry groups $SO(4)_L$ and $SO(4)_R$, respectively.
While looking at these superfields from the perspective of
the diagonal subgroup $SU(2)_{diag}$ in the product
$SU(2)_L\times SU(2)_R$, with the $SU(2)$ factors
being realized on the indices $i$ and $a$, these four types
of twisted multiplets amount, respectively,
to a sum of the $SU(2)_{diag}$ singlet and triplet superfields,
two complex doublet superfields and a sum of
four $SU(2)_{diag}$ singlet superfields. These sets provide
an off-shell extension of what was called
scalar multiplets SM-II, SM-III, SM-IV and SM-I in \cite{GRana,GKet}.
Two of these superfields, $\hat q^{\,i\,a}$ and $\hat q^{\,\und i\, \und a}$\,,
comprise just what was termed TM-II and TM-I twisted multiplets in \cite{G1,GKet}.
In \cite{GI,G1,GKet}, the superfield kinetic actions
were written for such multiplets, as well as
the invariant mass (potential) terms, and it was observed (see also \cite{IKL})
that the mixed mass terms can be composed only of the two multiplets
``dual'' to each other. In our notation, such ``self-dual'' pairs are formed by
the first and fourth, or second and third superfields from the above set.
The natural question is as to what is the most general self-interaction of
these four different species of twisted multiplets, both in regard
the sigma-model type of it (generalizing free kinetic terms) and
superpotential type (generalizing the mass terms). In both ${\cal N}=(2,2)$ \cite{GHR}
and ${\cal N}=(4,4)$ \cite{IS} superspace approaches only the general Lagrangians of
{\it one kind} of twisted multiplet were considered and the appropriate
restrictions on the relevant bosonic target metric and torsion were deduced.

One of the purposes of the present paper is to answer the question
just mentioned, using the bi-harmonic $SU(2)\times SU(2)$ approach
of refs. \cite{IS,IS1,bi1}. As a prerequisite, we give how these
four different twisted multiplets are described within this
setting. Only one of them (just the one comprised by the
superfield $\hat q^{\,i\,a}$ from the above set) is presented by an
analytic bi-harmonic superfield, and it is just the multiplet the
general ${\cal N}=(4,4)$ actions for which were given in \cite{IS}. The
remaining three multiplets have a more complicated description. We
firstly construct the general invariant superfield sigma-model
type actions for these multiplets, including an example of the
superconformal action which is the appropriate ${\cal N}=(4,4)$
superextension of $SU(2)\times U(1)$ WZNW model. Then we study the
mixed case when multiplets of different kind could interact with
each other. We find that ${\cal N}=(4,4)$ supersymmetry requires the
corresponding actions to split into a sum of actions for separate
multiplets, and this phenomenon is one of the basic findings of
our paper. Another one concerns the structure of admissible
superpotential terms. We find that such terms can be constructed
for each separate multiplet and/or for a pair of multiplets
``dual'' to each other. These
additional terms are defined in a unique way, and the form of the
corresponding component potential is uniquely fixed by the bosonic
target metric, like in the cases considered in \cite{IS,GI}.
Once again, ${\cal N}=(4,4)$ supersymmetry forbids possible
superpotential terms composed of the multiplets belonging to
different ``self-dual'' pairs. Thus these pairs cannot ``talk'' to
each other at all.

The paper is organized as follows. We start in Sect. 2 with recollecting
the basic facts about the $SU(2) \times SU(2)$ HSS and off-shell description
of the twisted analytic $q^{1,1}$ multiplet in its framework (corresponding
to the superfield $\hat q^{\,i\,a}$ from the set \p{base}). We also recall
the realization of ${\cal N}=4$, $2D$ superconformal groups in the analytic subspace
of this HHS. In Sect. 3 we give the description of the remaining three
twisted multiplets in $SU(2)\times SU(2)$ HSS and show that in
the analytic subspace they are presented by some analytic superfunctions
having nontrivial transformation properties under the supersymmetry.
Due to the latter circumstance, the supersymmetric actions of these multiplets
in the analytic subspace are written through the Lagrangians subjected to some
differential constraints required by supersymmetry, and these actions are
invariant up to a shift of the Lagrangians by a total derivative.
The relevant component actions are shown to be completely specified by
the metric on the physical bosons manifold. They reveal the same target geometry
as in the $q^{1,1}$ case \cite{IS}. In Sect. 4, on the example of the multiplet
$\hat q^{\,i\,\und a}$\,, we show that the requirement of invariance under one
of the four ``small'' $SU(2)$ superconformal groups which can be defined in the analytic
subspace uniquely specifies the relevant action to be that of ${\cal N}=(4,4)$ extension
of the group manifold $SU(2)\times U(1)$ WZNW sigma model. In Sect. 5 we construct
massive extensions of the superfield sigma model actions for two separate
multiplets and show that these extra terms are uniquely fixed by supersymmetry,
like in the $q^{1,1}$ case \cite{IS}. Sect. 6 is devoted to possible mixed
interactions of different multiplets. A careful analysis shows that the general
sigma model actions always split into a sum of actions for separate multiplets,
while the crossing-interaction through the mass terms is possible only for
twisted multiplets belonging to the same ``self-dual'' pair. We present the most
general form of the component potential term for such a pair arising as a result
of elimination of the auxiliary fields in the full action. This potential gets
contributions from the three sources: mass terms for each separate multiplet
and the mixed mass term.

\setcounter{equation}{0}
\section{$SU(2)\times SU(2)$ harmonic superspace}

We begin by recalling basics of ${\cal N}=(4,4),$ $2D$ supersymmetry.
The standard real ${\cal N}=(4,4),$ $2D$ superspace is parametrized
by the following set of the light-cone coordinates
$$
{\bf R}^{(1,1|4,4)} = (\,Z\,) =
(\,z^{++}\,, z^{--}\,, \theta^{+i \und k}\,, \theta^{-a \und b}\,)\,.
$$
Here $+$,$-$ are light-cone indices and $i$, $\und k$, $a$, $\und b$
are doublet indices of four commuting $SU(2)$ groups which constitute
the full automorphism group $SO(4)_L\times SO(4)_R$ of ${\cal N}=(4,4),$ $2D$
Poincar\'e superalgebra. The corresponding covariant spinor derivatives
obey the following algebra
\be
\{\,D_{i \und k}\,, D_{j \und l} \,\} = 2i\, \eps_{i\,j}\, \eps_{\und k\, \und l}\,
\partial_{++}\,, \quad
\{\,D_{a \und b}\,, D_{c \und d} \,\} = 2i\, \eps_{a\,c}\, \eps_{\und b\, \und d}\,
\partial_{--}
\label{alg}
\ee
where
\bea
D_{i\und k} = \frac{\partial}{\partial \theta^{i\und k}} +
i\theta_{i \und k}\,\partial_{++}\,, \qquad
D_{a \und b} = \frac{\partial}{\partial \theta^{a\und b}} +
i\theta_{a \und b}\,\partial_{--}
\eea
(hereafter, we omit the light-cone indices of the Grassmann coordinates,
keeping in mind the rule that the doublet indices $i, \und k$ refer to
the left sector, while $a, \und b$ to the right one).
Here we use the quartet notation for spinor derivatives and Grassmann
coordinates. Its relation to the complex notation of ref. \cite{IS} is as follows
\bea
\label{quart}
\theta^{i \und k} \equiv (\theta^i, \bar \theta^i)\,, \quad
D_{i \und k} \equiv (D_i\,, \bar D_i) \,, \quad
\theta^{a \und b} \equiv (\theta^a, \bar \theta^a) \,, \quad
D_{a \und b} \equiv (D_a\,, \bar D_a) \,.
\eea
The complex conjugation rules are
\be
(\theta^{i \und k})^{\dagger} =
\eps_{i\,l}\, \eps_{\und k\, \und n}\, \theta^{l \und n}\,,\quad
(D_{i \und k})^{\dagger} =
- \eps^{i\,l}\, \eps^{\und k\, \und n}\, D_{l \und n} \nn
\ee
(and the same for the objects from the right sector).

The ${\cal N}=(4,4)$ $SU(2)\times SU(2)$ harmonic superspace (HSS) introduced in
\cite{IS} is an extension of the real $2D$ superspace defined above by
two independent sets of harmonic variables $u^{\pm 1}_i$ and $v^{\pm 1}_a$
associated with one of the $SU(2)$ factors of the $SO(4)_L$ and $SO(4)_R$
automorphism groups of the left and right sectors of ${\cal N}=(4,4)$ supersymmetry,
respectively (we denote them by $SU(2)_L$ and $SU(2)_R$, this choice of
$SU(2)$ subgroups is optional). The $SU(2)\times SU(2)$ HSS formalism
enables one to keep both these $SU(2)$ symmetries manifest at each step
and to control their breakdown. We define the central basis of this HSS as
\be
{\bf HR}^{(1+2,1+2|4,4)} = (Z\,,u\,,v) =
{\bf R}^{(1,1|4,4)} \otimes (u^{\pm1}_i, v^{\pm1}_a)\,, \quad
u^{1i} u^{-1}_i = 1\,,\;\;\; v^{1a} v^{-1} _a = 1\,.
\label{HSS}
\ee

The analytic basis in the same ${\cal N}=(4,4)$ $SU(2)\times SU(2)$ HSS amounts
to the following choice of coordinates
\be
{\bf HR}^{(1+2,1+2|4,4)} = (X\,,u\,,v) = (\,x^{++}\,, x^{--}\,,
\theta^{\pm1,0\; \und i}\,, \theta^{0,\pm1\; \und a}\,,
u^{\pm 1}_i\,, v^{\pm 1}_a\,)
\label{an.set}
\ee
where
$$
\theta^{\pm 1,0\; \und i} = \theta^{k \und i}\, u^{\pm 1}_k\,, \quad
\theta^{0,\pm 1\; \und a} = \theta^{b \und a}\, v^{\pm 1}_b\,.
$$
The precise relation between $x^{\pm \pm}$ and $z^{\pm \pm}$ can
be found in \cite{IS}. The main feature of the analytic basis is
that it visualizes the existence of the {\it analytic subspace} in
the $SU(2)\times SU(2)$ HSS:
\be
{\bf AR}^{(1+2,1+2|2,2)} = (\zeta,u,v) =
(\,x^{++}\,, x^{--}\,, \ti\,, \ta\,, u^{\pm 1}_i\,, v^{\pm 1}_a\,)\,,
\label{AS}
\ee
which is closed under the ${\cal N}=(4,4)$
supersymmetry transformations. The existence of the analytic
subspace matches with the form of covariant spinor derivatives in
the analytic basis
\be
D^{1,0\; \und i} = -\frac{\partial}{\partial \theta^{-1,0}_{\und i}}\,,\quad
D^{0,1\; \und a} = -\frac{\partial}{\partial \theta^{0,-1}_{\und a}}
\label{sp.der}
\ee
where
\be
D^{\pm 1,0\; \und i} \equiv D^{k \und i}\, u^{\pm 1}_k\,,\quad
D^{0,\pm 1\; \und a} \equiv D^{b \und a}\, v^{\pm 1}_b\,.
\ee
The ``shortness'' of $D^{1,0 \;\und i}\,,
D^{0,1\; \und a}$ means that the Grassmann-analytic bi-harmonic
superfields $\Phi^{\,q,\,p}$,
\be
D^{1,0\; \und i}\, \Phi^{\,q,\,p} = D^{0,1\; \und a}\, \Phi^{\,q,\,p} = 0\,,
\label{Phi}
\ee
do not depend on
$\theta^{-1,0\; \und i}\,, \theta^{0, -1\; \und a}$ in the analytic
basis, i.e. are defined on the analytic superspace \p{AS}:
\be
\Phi^{\,q,\,p} = \Phi^{\,q,\,p} (\zeta, u, v)\,.
\ee

The pair of superscripts `$q,p$' on $\Phi^{\,q,\,p}$ in (\ref{Phi}),
as well as analogous superscripts on other quantities, stands for
the values of two independent harmonic $U(1)$ charges which, as in
the case of $SU(2)$ HSS \cite{HSS,book}, are assumed to be
strictly preserved. As a consequence of this requirement, all
superfields (or superfunctions), defined on (\ref{AS}), i.e. the
$SU(2)\times SU(2)$ {\it analytic} ${\cal N}=(4,4)$ superfields (or
superfunctions), are assumed to admit expansions in the double
harmonic series on the product of two 2-spheres
$SU(2)_L/U(1)_L\times SU(2)_R/U(1)_R$\,. The extra doublet indices
$\und i,\; \und a$ of Grassmann coordinates in (\ref{AS}) refer to
two additional $SU(2)$ automorphism groups of ${\cal N}=(4,4),$ $2D$
Poincar\'e supersymmetry which, together with $SU(2)_L$ and
$SU(2)_R$\,, constitute the full automorphism group $SO(4)_L\times
SO(4)_R$ of the latter. We prefer not to ``harmonize'' these
additional $SU(2)$ groups in order to avoid unwanted complications
in the notation.\footnote{${\cal N}=(4,4),$ $2D$ HSS with three sets of
$SU(2)$ harmonics was considered in \cite{Zupn}.}

In the bi-harmonic superspace one can define two sets of mutually
commuting harmonic derivatives, the left and right ones, each
forming an $SU(2)$ algebra \cite{IS}. Here we will need to know
the explicit expressions only for the derivatives with positive
$U(1)$ charges which commute with $D^{1,0\; \und i}\,, D^{0,1\; \und
a}$ and so preserve the harmonic analyticity. In the analytic
basis, these derivatives read
\be
\bigtriangledown^{2,0} =
\da + \ti\, \frac{\partial}{\partial \theta^{-1,0\; \und i}}\,, \quad
\bigtriangledown^{0,2} =
\db + \ta\, \frac{\partial}{\partial \theta^{0,-1\; \und a}}
\label{Harm.der}
\ee
where
\be
\da =
\partial^{2,0} + i \ti\, \theta^{1,0}_{\und i} \partial_{++}\,,\quad
\db = \partial^{0,2} + i \ta\, \theta^{0,1}_{\und a}
\partial_{--} \label{harm.der}
\ee
and
$$
\partial^{2,0} = u^{1i} \frac{\partial}{\partial u^{-1i}}\,, \quad
\partial^{0,2} = v^{1a} \frac{\partial}{\partial v^{-1a}}\,.
$$
When acting on the analytic superfields, $\bigtriangledown^{2,0}$
and $\bigtriangledown^{0,2}$ are reduced to $\da$ and $\db$\,.

The main advantage of using the $SU(2)\times SU(2)$ HSS (as compared e.g.
with the standard $SU(2),$ $2D$ HSS obtained as a dimensional reduction
of ${\cal N}=2,$ $4D$ HSS \cite{HSS,book}) consists in the fact
that it provides a natural superfield description for the
important class of ${\cal N}=(4,4)$ supersymmetric sigma models, those with
torsion on the bosonic manifold, such that the whole amount of
the underlying ${\cal N}=(4,4)$ supersymmetry is manifest and off-shell.

Here we recall the HSS off-shell formulation of the first type of
${\cal N}=(4,4)$ twisted multiplets from the set \p{base}, postponing the HSS
treatment of the remaining three types to the next Section.
This multiplet was used in \cite{IS} to construct the ${\cal N}=(4,4)$ superspace
version of the general ${\cal N}=(2,2)$ superspace action of one sort of ${\cal N}=(4,4)$
twisted multiplet \cite{GHR}.
In HSS this multiplet is described by a real analytic ${\cal N}=(4,4)$ superfield
$q^{1,1}(\zeta, u, v)$ subjected to the harmonic constraints
\be
\da q^{1,1} = 0\,,\quad \db q^{1,1} = 0\,.
\label{hc1}
\ee
These constraints leave $(8+8)$ independent components in $q^{1,1}$  \cite{IS},
which is just the irreducible {\it off-shell} component content of ${\cal N}=(4,4)$
twisted multiplet. In the central basis the constraints \p{hc1} and
the analyticity conditions imply
\be
q^{1,1} = \hat q^{\,i\,a}u^{1}_i v^{1}_a\,,
\quad D^{(k \und k}\hat q^{\,i)\, a} = D^{(b \und b}\hat q^{\,k\, a)} = 0\,,
\label{q11constr}
\ee
and we end up with the first type of twisted multiplet from the set \p{base}
(the form of constraints as in \p{q11constr} was exhibited for the first time in \cite{IK}).
The analytic basis solution of the harmonic constraints \p{hc1} is given by
\bea
q^{1,1}&=& q^{i\,a} u^1_i v^1_a  + \ti \varphi^a_{\und i} v^1_a
+ \ta \eta^i_{\und a} u^1_i -
i (\theta^{1,0})^2 \partial_{++} q^{i\,a} u^{-1}_i v^1_a
-i (\theta^{0,1})^2 \partial_{--} q^{i\,a} u^1_i v^{-1}_a \nn\\
&+& \ti \ta F_{\und i\; \und a} - i \ti (\theta^{0,1})^2
\partial_{--} \varphi^a_{\und i} v^{-1}_a - i (\theta^{1,0})^2 \ta
\partial_{++} \eta^i_{\und a} u^{-1}_i  \nn \\
&-& (\theta^{1,0})^2 (\theta^{0,1})^2
\partial_{++} \partial_{--} q^{i\,a} u^{-1}_i v^{-1}_a\,
\label{567}
\eea
where $(\theta^{1,0})^2 = \theta^{1,0\; \und k}\,\theta^{1,0}_{\und k}$\,,
$(\theta^{0,1})^2 = \theta^{0,1\; \und a}\,\theta^{0,1}_{\und a}$\,.

The general {\it off-shell} action of $n$ such superfields $q^{1,1\, M}$
$(M = 1,2,...n)$ can be written as the following integral over the analytic
superspace (\ref{AS})
\be
S^{gen} = \int \mu^{-2,-2}\, {\cal L}^{2,2} (q^{1,1\, M},u,v)
\label{s1gen}
\ee
where
\be
\mu^{-2,-2} = d^2x\,d^2\theta^{1,0}\, d^2\theta^{0,1}\, du\, dv
\label{measure}
\ee
is the analytic superspace integration measure (see Appendix for its precise definition).
In general, the dimensionless analytic superfield Lagrangian ${\cal L}^{2,2}$ bears an arbitrary
dependence on its arguments, the only restriction being a compatibility
with its external $U(1)$ charges $(2,2)$\,. The free action is given by
\be
S^{free} \sim \int \mu^{-2,-2}\, q^{1,1\, M} q^{1,1\, M}\,,
\ee
so for consistency we are led to assume
$$
\left. \det\left(\frac{\partial^2 {\cal L}^{2,2}}
{\partial q^{1,1\, M}\,\partial q^{1,1\, N}}\right) \right |_{q^{1,1} = 0} \ne 0\,.
$$
The passing to the component form of the action (\ref{s1gen}) is straightforward
\cite{IS}. The relevant bosonic sigma model action consists of two parts related to
each other by ${\cal N}=(4,4)$ supersymmetry: the metric part and the part which includes
the torsion potential. These terms obey the same constraints as in the ${\cal N}=(2,2)$
superspace description of general torsionful sigma models associated with
twisted ${\cal N}=(4,4)$ multiplets of one type \cite{GHR}. So \p{s1gen} provides
a manifestly ${\cal N}=(4,4)$ supersymmetric form of the general action of such multiplets.

As the last topic of this Section we remind some details of the
${\cal N}=4$, $2D$ superconformal groups. As discussed in \cite{IS}, in the
$SU(2)\times SU(2)$ analytic HSS one can realize two different infinite-parameter
``small' ${\cal N}=4$, $SU(2)$ superconformal groups (in each light-cone sector),
having as their closure the ``large'' ${\cal N}=4$, $SO(4)\times U(1)$ superconformal group.
One of these ${\cal N}=4$, $SU(2)$ superconformal groups acts on all coordinates
of the analytic HSS, including the harmonic coordinates \cite{IS,DS}:
\bea
\delta x^{++} = \tilde \Lambda_{(I)L}\,, \;
\delta u^1_i = \Lambda^{2,0} u^{-1}_i\,,\;
\delta u^{-1}_i = 0\,,\;
\delta \ti = \Lambda_{(I)}^{1,0\; \und i}\,, \;
\delta \da = -\Lambda^{2,0} \du\,,
\label{tran1L}
\eea
\bea
\delta x^{--} = \tilde \Lambda_{(I)R}\,, \;
\delta v^1_a = \Lambda^{0,2} v^{-1}_a\,,\;
\delta v^{-1}_a = 0\,,\;
\delta \ta = \Lambda_{(I)}^{0,1\; \und a}\,,\;
\delta \db = -\Lambda^{0,2} \dv\,.
\label{tran1R}
\eea
Here $\du$ and $\dv$ are the left and right $U(1)$ charge-counting operators
and
\bea
\label{sc.sol}
&& \tilde \Lambda_{(I)L} = a_L -\frac{1}{2}\, \partial^{-2,0} \da a_L\,,\;\;
\Lambda^{2,0} = \da \Lambda_{(I)L}\,,\;\;
\Lambda_{(I)L} = -\frac{1}{2}\, \partial_{++} a_L\,,\;\; \nn\\
&&
\Lambda_{(I)}^{1,0\; \und i} = - \eps^{\und i\, \und k}\, \frac{i}{4}
\,\frac{\partial}{\partial \theta^{1,0\; \und k}}\, \da a_L\,, \;\;
(\da)^2\, a_L = 0\,,
\eea
\bea
&&
\tilde \Lambda_{(I)R} = a_R -\frac{1}{2}\, \partial^{0,-2} \db a_R\,,\;\;
\Lambda^{0,2} = \db \Lambda_{(I)R}\,,\;\;
\Lambda_{(I)R} = -\frac{1}{2}\, \partial_{--} a_R\,,\;\;  \nn\\
&&
\Lambda_{(I)}^{0,1\; \und a} = - \eps^{\und a\, \und b}\, \frac{i}{4}
\,\frac{\partial}{\partial \theta^{0,1\; \und b}}\, \db a_R\,, \;\;
(\db)^2\, a_R = 0\,.
\eea
The superparameter functions $a_L$ and $a_R$ depend only on the left and right
light-cone coordinates, respectively, i.e. $a_L = a_L(\zeta_L,u)$ and
$a_R = a_R (\zeta_R, v)$ where $\zeta_L = (\,x^{++}\,, \theta^{1,0\;\und i}\,), \;\zeta_R
= (\,x^{--}\,, \theta^{0,1\;\und a}\,)\,.$
The explicit form of these functions can be found in \cite{IS}.
In what follows we shall need the identities
\be
\Lambda_{(I)}^{1,0\; \und i} = \da \Lambda_{(I)}^{-1,0\; \und i} - \ti \Lambda_{(I)L}\,,
\quad
\Lambda_{(I)}^{0,1\; \und a} = \db \Lambda_{(I)}^{0,-1\; \und a} - \ta \Lambda_{(I)R}
\label{iden1}
\ee
where
\be
\Lambda_{(I)}^{-1,0\; \und i} = - \eps^{\und i\, \und k}\, \frac{i}{4}
\,\frac{\partial a_L}{\partial \theta^{1,0\; \und k}}\,,\quad
\Lambda_{(I)}^{0,-1\; \und a} = - \eps^{\und a\, \und b}\, \frac{i}{4}
\,\frac{\partial a_R}{\partial \tb}\,.
\ee
They can be proved using \p{sc.sol}.

Another ${\cal N}=4$, $SU(2)$ superconformal group (also consisting of two
mutually commuting left and right components)
does not affect harmonic variables
\bea
\label{tran2LR}
&&
\delta x^{++} = \Lambda_{(II)L}\,,\quad
\delta \ti = \Lambda_{(II)}^{1,0\; \und i}\,,\quad
\delta u^{\pm 1}_i = 0\,,\nn\\
&&
\delta x^{--} = \Lambda_{(II)R}\,,\quad
\delta \ta = \Lambda_{(II)}^{0,1\; \und a}\,,\quad
\delta v^{\pm 1}_a = 0
\eea
and is fully specified by requiring $\da$, $\db$ to be invariant
\be
\delta \da = 0\,,\qquad  \delta \db = 0\,.
\ee
The latter equations imply
\bea
&&
\da \Lambda_{(II)}^{1,0\; \und i} = 0\,, \quad
\da \Lambda_{(II)L} =
2i\, \theta_{\und i}^{1,0} \Lambda_{(II)}^{1,0\; \und i}\,, \nn\\
&&
\db \Lambda_{(II)}^{0,1\; \und a} = 0\,, \quad
\db \Lambda_{(II)R} =
2i\, \theta_{\und a}^{0,1} \Lambda_{(II)}^{0,1\; \und a}\,.
\label{a}
\eea
The general solution to eqs. (\ref{a}) is provided by
\bea
&&
\Lambda_{(II)}^{1,0\; \und i} = \lambda^{k\, \und i}u^1_k + \tk\,
(\lambda^{\und i\,)}_{(\, \und k} - \frac{1}{2}\, \delta^{\und i}_{\und k}\,
\partial_{++} \lambda_L)
- i\,(\theta^{1,0})^2 \partial_{++} \lambda^{k\, \und i} u^{-1}_k\,, \nn \\
&&
\Lambda_{(II)L} = \lambda_L + 2i\, \theta_{\und i}^{1,0} \lambda^{k\, \und i} u^{-1}_k\,, \nn\\
&&
\Lambda_{(II)}^{0,1\; \und a} = \lambda^{b\, \und a}v^1_b + \tb\,
(\lambda^{\und a\,)}_{(\, \und b} - \frac{1}{2}\, \delta^{\und a}_{\und b}\,
\partial_{--} \lambda_R)
- i\,(\theta^{0,1})^2 \partial_{--} \lambda^{b\, \und a} v^{-1}_b \,, \nn \\
&&
\Lambda_{(II)R} = \lambda_R +2i\, \theta_{\und a}^{0,1} \lambda^{b\, \und a} v^{-1}_b\,.
\label{sc.sol2}
\eea
We also quote the identities to be used in what follows:
\be
\da \Lambda_{(II)}^{-1,0\; \und i} = \Lambda_{(II)}^{1,0\; \und i}
- \tk\, \frac{\partial \Lambda_{(II)}^{1,0\; \und i}}{\tk}\,, \quad
\db \Lambda_{(II)}^{0,-1\; \und a} = \Lambda_{(II)}^{0,1\; \und a}
- \tb\, \frac{\partial \Lambda_{(II)}^{0,1\; \und a}}{\tb}
\label{iden2}
\ee
where
\be
\Lambda_{(II)}^{-1,0\; \und i} \equiv \lambda^{k\, \und i} u^{-1}_k\,, \quad
\Lambda_{(II)}^{0,-1\; \und a} \equiv \lambda^{b\, \und a} v^{-1}_b\,.
\ee
They can be proved using the explicit expressions \p{sc.sol2}.

The analytic superfield $q^{1,1}$ defined by eqs. \p{hc1} behaves as a scalar
of conformal weight zero under the action of the ${\cal N}=4$, $SU(2)$ superconformal
group $II$, but possesses nontrivial transformation properties
under the ${\cal N}=4$, $SU(2)$ superconformal group $I$ \cite{IS}:
\be
\delta_{II}\, q^{1,1} = 0\,, \quad \delta_I\, q^{1,1} = \Lambda_{(I)L}\, q^{1,1}
\label{SCq11}
\ee
(the transformation rules with respect to the right branches of these
superconformal groups are the same up to the change $L\rightarrow R$).

In the realization on $q^{1,1}$\,, the basic difference between the superconformal
groups $I$ and $II$ manifests itself in the action of their $SU(2)$ subgroups.
The left and right $SU(2)$'s belonging to the superconformal group $I$ act on
the indices $i$ and $a$ and so possess a nontrivial action on both the physical
bosons $q^{i\,a} = \hat q^{\,i\,a}|$ and the fermions; at the same time,
$SU(2)$ subgroups from the superconformal group $II$ act on the indices
$\und i, \und a$ and so affect only fermions and auxiliary fields in $q^{1,1}$\,.
Note that the above pairing of two left and two right ${\cal N}=4$, $SU(2)$ superconformal
groups into the ${\cal N}=(4,4)$ superconformal groups $I$ and $II$ is optional:
one could alternatively assemble one of such ${\cal N}=(4,4)$ supergroups as a direct
product of the left branch of the superconformal group $I$ and the right branch of
the superconformal group $II$, and the second ${\cal N}=(4,4)$ superconformal group
as the product of the right branch of $I$ and the left branch of $II$.
These different possibilities of composing ${\cal N}=(4,4)$ superconformal groups out of
the mutually commuting left and right pairs of ${\cal N}=4$, $SU(2)$ superconformal groups
are directly related to the existence of four different sorts of twisted multiplets,
as given in \p{base} and explained in more details in next Sections. For every of
these multiplets one can single out the proper ${\cal N}=(4,4)$ superconformal groups which
act on them precisely in the same way as the  superconformal groups $I$
and $II$ defined above act on the multiplet $\hat q^{\,i\,a} \sim q^{1,1}$\,.

Finally, we wish to mention that the analytic superspace integration measure
\p{measure} is invariant with respect to both superconformal groups \cite{IS}.

\setcounter{equation}{0}

\section{New types of twisted multiplets in $SU(2)\times SU(2)$ HSS}

\subsection{Constraints in the general and analytic superspaces}

As mentioned in Introduction, in the standard
real ${\cal N}=(4,4),$ $2D$ superspace we can define four types of
twisted multiplets in accord with the four possibilities to pair the doublet
indices of various $SU(2)$ factors of the left and right R-symmetry
groups $SO(4)_L$ and $SO(4)_R$\,. The first type is $\hat q^{\,i\,a}$ studied
in \cite{IS}. Its HSS description was reminded in the previous Section.
The remaining three ones are
\bea
(\mbox{a})\;\; \hat q^{\,i\, \und a}\,, \quad (\mbox{b}) \;\; \hat q^{\,\und i\, a}\,,
\quad (\mbox{c})\;\; \hat q^{\,\und i\, \und a}\,,\label{base1}
\eea
\bea
(\hat q^{\,i\, \und a})^\dagger =
\eps_{i\,k}\, \eps_{\und a \,\und b}\, \hat q^{\,k\, \und b}\,,\quad
(\hat q^{\,\und i\, a})^\dagger =
\eps_{\und i \,\und k}\, \eps_{a\, b}\, \hat q^{\,\und k\, b}\,,\quad
(\hat q^{\,\und i\, \und a})^\dagger = \eps_{\und i \,\und k}\,
\eps_{\und a \,\und b}\, \hat q^{\,\und k\, \und b}\,. \nn
\eea
The real quartet superfields in \p{base1} are subjected to the irreducibility
conditions which are quite similar to those defining $\hat q^{\,i\,a}$
in the central basis (see eq. \p{q11constr})
\bea
&&
(\mbox{a})\;\; D^{(k \und k} \hat q^{\,i)\, \und a} =
D^{b (\und b} \hat q^{\,i\, \und a)} = 0\,, \quad
(\mbox{b})\;\; D^{k (\und k} \hat q^{\,\und i)\, a} =
D^{(b \und b} \hat q^{\,\und i \, a)} = 0\,, \nn\\
&&
(\mbox{c}) \;\; D^{k (\und k} \hat q^{\,\und i)\, \und a} =
D^{b (\und b} \hat q^{\,\und i\, \und a)} = 0\,.
\label{stan.con.2}
\eea
Clearly, these constraints like \p{q11constr} imply that
all superfields \p{base1} carry out the same
off-shell content $(8+8)$, though with a different assignment of the component
fields with respect to four $SU(2)$ groups which form $SO(4)_L\times SO(4)_R$\,.

Converting the $SU(2)$ indices of the superfields $\hat q^{\,i\, \und a}$\,,
$\hat q^{\,\und i\,a}$\,, $\hat q^{\,\und i\, \und a}$ and spinor derivatives
in (\ref{stan.con.2}) with the harmonics $u_i^1, v_a^1$, we can rewrite these
constriants in the analytic basis (\ref{an.set}) of the $SU(2)\times SU(2)$ HSS as
\bea
&& (\mbox{a}) \;\; D^{1,0 \; \und k}\, \hat q^{\,1,0 \; \und a} =
D^{0,1 \, (\und b}\, \hat q^{\,1,0 \; \und a)} = 0\,,
\quad
(\mbox{b}) \;\; D^{1,0\, (\und k}\, \hat q^{\,0,1\; \und i)} =
D^{0,1\; \und b}\, \hat q^{\,0,1\; \und i} = 0\,, \nn \\
&& (\mbox{c}) \;\; D^{1,0\,(\und k}\, \hat q^{\,\und i)\, \und a}
= D^{0,1\,(\und b}\, \hat q^{\,\und i\, \und a)} = 0
\label{an.con.4}
\eea
where, in the central basis,
\be
\hat q^{\,1,0 \; \und a} = \hat q^{\,i\, \und a}u^1_i\,, \quad
\hat q^{\,0,1\; \und i} = \hat q^{\,\und i\, a}v^1_a\,.
\ee

Using the analytic basis form of $D^{1,0 \; \und i}\,,\, D^{0,1 \; \und a}$\,,
eq. (\ref{sp.der}), and expanding the superfields in the non-analytic
odd coordinates $\theta^{-1,0\; \und i}\,, \,\theta^{0,-1\; \und a}$\,,
one can solve (\ref{an.con.4}) in the analytic basis as
\bea
&&
\kern-12mm (\mbox{a})\;\;\hat q^{\,1,0 \; \und a}(X,u,v)
= \qa + \theta^{0,-1 \; \und a}\, g^{1,1}\,,
\quad (\mbox{b}) \;\; \hat q^{\,0,1\; \und i}(X,u,v)
= \qb + \theta^{-1,0 \; \und i}\, f^{1,1}\,,
\label{sol.3} \\
&&
\kern-12mm (\mbox{c}) \;\; \hat q^{\,\und i \,\und a}(X,u,v) =
\tilde{q}^{\,\und i \,\und a}
+ \theta^{-1,0 \;\und i}\, f^{1,0 \;\und a} +
\theta^{0,-1 \;\und a}\, h^{0,1\; \und i}
+ \theta^{-1,0\; \und i}\, \theta^{0,-1\; \und a}\, t^{1,1}
\label{sol.4}
\eea
where all the coefficients depend only on the analytic coordinates $(\zeta, u, v)$\,.

Thus the harmonic superfields $\hat q^{\,1,0 \; \und a}$\,, $\hat q^{\,0,1\; \und i}$\,,
$\hat q^{\,\und i \,\und a}$ bear the explicit dependence on the non-analytic
Grassmann coordinates with the negative $U(1)$ charge and so are not harmonic-analytic,
as opposed to the superfield $q^{1,1}$\,.
On the other hand, all the components in their expansions over the non-analytic
coordinates are defined on the analytic subspace of the $SU(2)\times SU(2)$ HSS.
The basic difference between them and the analytic superfield $q^{1,1}$
consists in their supersymmetry transformation properties. Keeping in mind that
$\hat q^{\,1,0 \; \und a}$\,, $\hat q^{\,0,1\; \und i}$\,, $\hat q^{\,\und i \,\und a}$
are the general harmonic superfields
(here $\delta$ is the variation under supersymmetry)
\be
\delta  \hat q^{\,1,0 \; \und a} = \delta \hat q^{\,0,1\; \und i} =
\delta  \hat q^{\,\und i \,\und a} = 0\,,
\ee
and the supertranslations of $\theta$'s are
$\delta \theta^{-1,0\; \und i} = \eps^{-1,0\; \und i}\,,\;
\delta \theta^{0,-1\; \und a} = \eps^{0,-1\; \und a}\,,$
it is easy to find how these analytic components are transformed
\bea
(\mbox{a}) \kern-4mm
&&\delta \qa = -\eps^{0,-1 \; \und a}\, g^{1,1}\,, \;\; \delta g^{1,1} = 0\,,
\quad  (\mbox{b})\;\; \delta \qb = -\eps^{-1,0 \; \und i}\, f^{1,1}\,, \;\;
\delta f^{1,1} = 0\,,
\label{tran3} \\
(\mbox{c}) \kern-4mm &&\delta  \tilde{q}^{\, \und i \,\und a} =
- \eps^{-1,0 \; \und i}\, f^{1,0 \; \und a} - \eps^{0,-1 \; \und a}\, h^{0,1 \; \und i}\,, \;\;
\delta  f^{1,0 \; \und a} = - \eps^{0,-1 \; \und a}\, t^{1,1}, \nn \\
&&
\delta h^{0,1 \; \und i} = - \eps^{-1,0 \; \und i}\, t^{1,1}, \;\;
\delta  t^{1,1} = 0\,.
\label{tran4}
\eea
Looking at (\ref{tran3}), (\ref{tran4}), one observes that the highest
components of $\hat q^{\,1,0 \; \und a}$\,, $\hat q^{\,0,1\; \und i}$\,,
$\hat q^{\,\und i \,\und a}$ are the genuine analytic superfields, while
$\qa$\,, $\qb$ and $\tilde{q}^{\, \und i \,\und a}$\,, despite being functions
of $(\zeta, u, v)$, are not analytic superfields in the rigorous sense since
they possess non-standard transformation properties under supersymmetry.
Note that $\qa$ and $\qb$ are still superfields of the left and right light-cone
Poincar\'e supersymmteries, respectively, while $\tilde q^{\,\und i \, \und a}$
possesses non-standard transformation properties under both supersymmetries.

Let us now re-express the central-basis property that the superfields
$\hat q^{\,1,0 \; \und a}$\,, $\hat q^{\,0,1\; \und i}$ and $\hat q^{\,\und i \,\und a}$
have the constrained dependence on the harmonics ($\hat q^{\,1,0 \; \und a}$ and
$\hat q^{\,0,1\; \und i}$ are linear in harmonics, while $\hat q^{\,\und i \,\und a}$
does not depend on them at all) as the following equivalent harmonic constraints
in the analytic basis
\be
\bigtriangledown^{2,0} \hat q^{\,1,0 \; \und a} =
\bigtriangledown^{0,2} \hat q^{\,1,0 \; \und a} =
\bigtriangledown^{2,0} \hat q^{\,0,1\; \und i} =
\bigtriangledown^{0,2} \hat q^{\,0,1\; \und i} =
\bigtriangledown^{2,0} \hat q^{\,\und i \,\und a} =
\bigtriangledown^{0,2} \hat q^{\,\und i \,\und a} = 0\,.
\label{harm.con}
\ee
Here, $\bigtriangledown^{2,0}$ and $\bigtriangledown^{0,2}$ are the full
analytic-basis harmonic derivatives defined in eqs. (\ref{Harm.der}),
(\ref{harm.der}). Substituting the expansions (\ref{sol.3}), (\ref{sol.4})
for $\hat q^{\,1,0 \; \und a}$\,, $\hat q^{\,0,1\; \und i}$\,,
$\hat q^{\,\und i \,\und a}$ into \p{harm.con}, we can rewrite the latter
in a more detailed form as
\bea
&& \kern-5mm
\da \qa = 0\,, \quad \db \qa + \ta\, g^{1,1} =0\,,\quad
\da g^{1,1} = \db g^{1,1} = 0\,,
\label{harm.con2}\\
&& \kern-5mm
\da \qb + \ti\, f^{1,1} = 0\,, \quad \db \qb = 0\,, \quad
\da f^{1,1} = \db f^{1,1} = 0\,,
\label{harm.con3}
\eea
\bea
&&
\da \tilde{q}^{\, \und i \,\und a} + \ti\, f^{1,0 \; \und a} = 0\,,\quad
\db \tilde{q}^{\,\und i \,\und a} + \ta\, h^{0,1 \; \und i} = 0\,,\nn \\
&&
\da f^{1,0 \; \und a} = 0\,, \quad
\db f^{1,0 \; \und a} + \ta\, t^{1,1} =0\,, \nn \\
&&
\da h^{0,1 \; \und i} - \ti\, t^{1,1} = 0\,, \quad
\db h^{0,1 \; \und i} = 0\,, \nn\\
&&
\da t^{1,1} =\db t^{1,1} = 0\,.
\label{harm.con4}
\eea
These constriants are solved by
\bea
\qa
&=& q^{i\, \und a} u^1_i + \ti \alpha^{\und a}_{\und i} + \ta
\beta^{i\,a} u^1_i v^{-1}_a + \ti \ta F^a_{\und i} v^{-1}_a \nn\\
&-& \, i(\theta^{1,0})^2 \partial_{++} q^{i\, \und a} u^{-1}_i
- i (\theta^{1,0})^2 \ta \partial_{++} \beta^{i\,a} u^{-1}_i v^{-1}_a\,,
\label{bos.con2}
\\[0.3cm]
\qb &=& q^{\und i\, a} v^1_a + \ti \rho^{i\,a} u^{-1}_i v^1_a + \ta
\gamma^{\und i}_{\und a} + \ti \ta F^i_{\und a} u^{-1}_i \nn\\
&-& \, i (\theta^{0,1})^2 \partial_{--} q^{\und i\, a} v^{-1}_a - i
\ti (\theta^{0,1})^2 \partial_{--} \rho^{i\,a} u^{-1}_i v^{-1}_a\,,
\label{bos.con3} \\[0.3cm]
\tilde{q}^{\, \und i \,\und a}
&=& q^{\und i \,\und a} - \ti \psi^{i\, \und a} u^{-1}_i
- \ta \xi^{\und i\, a} v^{-1}_a + \ti \ta F^{i\,a} u^{-1}_i v^{-1}_a
\label{bos.con4}
\eea
where all the coefficients are $2D$ fields, $q^{i\,\und a} = q^{i\,\und a}(x)$\,, etc.
The expressions for the remaining analytic components are given in Appendix.
It is important to realize that all off-shell fields are collected already in
$\qa\,,\, \qb$ and $\tilde{q}^{\, \und i \,\und a}$\,, while the remaining analytic
superfunctions contain no new fields. The component expansions
\p{bos.con2} - \p{bos.con4} are to be compared with that of $q^{1,1}$ obtained
by solving the harmonic constraints \p{hc1} in the analytic basis (eq. \p{567}).

We observe a sort of duality inside the pairs $(q^{1,1}\,,
\;\hat{q}^{\, \und i \,\und a})$ and $(\hat q^{\,1,0 \; \und a}\,,
\;\hat q^{\,0,1 \; \und i})$\,: the $SU(2)$ assignments of the physical and
auxiliary bosonic fields in the first and second superfields within each pair
are reversed with respect to each other, while the assignments of fermions are
the same. As we shall see later, only these mutually ``dual'' twisted multiplets
can interact (through the proper superpotential terms).

\subsection{General actions of the superfields $\hat q^{\,1,0 \; \und a}$
and $\hat q^{\, \und i\, \und a}$}

As given in Sect. 2, the most general action \p{s1gen} of $n$ twisted multiplets
carried out by the analytic superfields $q^{1,1\,M} (M=1, \ldots n)$ is written as
an analytic superspace integral of the Lagrangian which is a generic charge $(2,2)$
function of $q^{1,1\,M}$ and harmonic variables $u^{\pm 1}_i, v^{\pm 1}_a\,$.
Being an analytic superfield, such a Lagrangian is manifestly invariant under
the supersymmetry transformations. On the other hand, the analytic superfunctions
representing other three twisted multiplets are not the standard superfields,
therefore their functionals are not superfields as well. As a result, constructing
the general supersymmetric actions of these multiplets
in the analytic superspace is not so straightforward as in the case of $q^{1,1}$\,.

The only primary principles (besides reality) of such a construction are: {\it (i).}
The preservation of two harmonic $U(1)$ charges whence it follows that the relevant
Lagrangian density should have the $U(1)$ charges $(2,2)$ for the action to be chargeless;
{\it (ii).} $2D$ Lorentz covariance which implies the Lagrangian density
${\cal L}^{2,2}$ to be Lorentz singlet; {\it (iii).} Dimensionality reasoning
which imply ${\cal L}^{2,2}$ to have the ``engineering dimension'' zero
(i.e. the same as the superfields $\hat q$ in \p{base1} have).

After constructing a general ${\cal L}^{2, 2}$ obeying these criteria,
one should examine which additional constraints are to be imposed on it
for the action to be invariant under the transformations \p{tran3} and
 \p{tran4} (possibly, up to a shift of the Lagrangian by a total derivative).

Before turning to the general case of the actions which simultaneously contain
a few different twisted ${\cal N}=(4,4)$ multiplets, we consider the actions with
only one type of the non-standard twisted multiplets \p{base1} as proper
analogs of the $q^{1,1}$ action \p{s1gen}. Without loss of generality,
it suffices to examine such actions only for $\hat q^{\,i\, \und a}$ and
$\hat q^{\,\und i\, \und a}$ since the action for $\hat q^{\,\und i\, a}$
can be recovered from that for $\hat q^{\,i\, \und a}$ via simple substitutions
including the replacement $v^{\pm 1}_a \leftrightarrow u^{\pm 1}_i\,$.

We firstly consider the actions of single multiplets.

In accordance with the primary principles above, the most general candidate
actions of the superfields $\hat q^{\,1,0 \; \und a}$\,, $\hat q^{\,\und i\, \und a}$
can be chosen as the following
integrals over the analytic superspace
\be
S^{gen}_{(a)} = \int \mu^{-2,-2}\, {\cal L}^{2,2}_{(a)}\,(\qa,g^{1,1},\ta,u,v)\,,
\label{s2gen}
\ee
\be
S^{gen}_{(c)} = \int \mu^{-2,-2}\, {\cal L}^{2,2}_{(c)}\,(\tilde{q}^{\, \und i \,\und a},
f^{1,0\; \und a}, h^{0,1\; \und i}, t^{1,1},
\theta^{1,0\; \und i}, \theta^{0,1\; \und a},u,v)
\label{s3gen}
\ee
where $\mu^{-2,-2}$ is the analytic superspace integration measure
defined in \p{measure}. Note that the left Grassmann coordinates
$\theta^{1,0\;\und i}$ cannot explicitly appear in ${\cal L}^{2,2}_{(a)}$
since $q^{1,0\;\und a}$ and $g^{1,1}$ are superfields with respect to the left
${\cal N}=4$ supersymmetry. On the other hand, under the right supersymmtery
$q^{1,0\;\und a}$ has non-standard transformation properties (see (\ref{tran3}a)),
and this is the reason why $\ta$ is included as a possible explicit argument in
${\cal L}^{2,2}_{(a)}$\,. Both types of Grassmann coordinates are admissible as
explicit arguments in ${\cal L}^{2,2}_{(c)}$ since $\tilde{q}^{\,\und i \,\und a}\,,
f^{1,0\; \und a}$ and $h^{0,1\; \und i}$ possess non-standard transformation
properties with respect to both left and right supersymmetries.

In order to further specify the Lagrangians in (\ref{s2gen}), (\ref{s3gen}),
we can resort to the following reasoning.
First, we must require that all possible terms in them are Lorentz invariant.
Second, we can rule out the dependence on all involved superfields
and superfunctions in (\ref{s2gen}), except for $\qa\,$ and harmonics,
and in (\ref{s3gen}), except for $\tilde{q}^{\,\und i \,\und a}$ and harmonics.
Indeed, Lorentz invariance requires that e.g. $g^{1,1}$ could enter (\ref{s2gen})
only as $\ta\, g^{1,1}$ (also taking into account that the fermionic superfield
$g^{1,1}$ is nilpotent and, hence, its any degree is vanishing). Then the harmonic
constraints (\ref{harm.con2}) imply that such a term can be written as $\db \qa$\,.

Taking into account this reasoning (and a similar one for the
$\hat q^{\,\und i \,\und a}$ multiplet), we can cast the Lagrangians in (\ref{s2gen}), (\ref{s3gen})
in the following more detailed form
\bea
S^{gen}_{(a)} &=& \int \mu^{-2,-2}\, \{\,{\cal L}^{2,2}_0(\qa,u,v)
+ {\cal L}^{1,0}_{\und a}\,(\qa,u,v)\, \db \qa \nn\\
&+& (\theta^{0,1})^2
\hat {{\cal L}}^{1,0}_{\und a}\,(\qa,u,v)\, \partial_{--} \qa \,\}\,,
\label{lagr.gen2}
\eea
\bea
S^{gen}_{(c)} &=& - \int \mu^{-2,-2}\, \{\, \hat{\cal L}^{2,2}_0(\tilde{q}^{\,\und i \,\und a},u,v) +
{\cal L}_{\und i\; \und a}^{2,0}\, (\tilde{q}^{\,\und i \,\und a},u,v)\, \db \tilde{q}^{\,\und i \,\und a}  \nn\\
&+& {\cal L}_{\und i\; \und a}^{0,2}\,(\tilde{q}^{\,\und i \,\und a},u,v)\, \da \tilde{q}^{\,\und i \,\und a}
+ {\cal L}_{\und i\; \und a\; \und k\; \und b}\,(\tilde{q}^{\,\und i \,\und a},u,v)\,
\da \tilde{q}^{\,\und i \,\und a}\, \db \tilde{q}^{\,\und k\, \und b}\, \}\,.
\label{lagr.gen3}
\eea
The reason why possible terms with $x$-derivatives are not included
in (\ref{lagr.gen3}) will become clear soon.

Now we turn to discussing the properties of the Lagrangians
${\cal L}^{2,2}_{(a)}$ and ${\cal L}^{2,2}_{(c)}$ under supersymmetry.

We start with (\ref{lagr.gen2}). As we know, $\qa$ is not a superfield.
It has a nontrivial transformation law \p{tran3} under the supersymmetry.
So we need to find which constraints should be imposed on the functions
in the r.h.s. of (\ref{lagr.gen2}) for ensuring the latter to be invariant.
The requirement of invariance amounts to the condition that the variation
of (\ref{lagr.gen2}) under the supersymmetry transformations is a sum of
total derivatives of arbitrary functions,
\bea
\delta \, {\cal L}^{2,2}_{(a)} = \db(\eps^{0,-1\, \und a}\,F^{\,2,1}_{\und a})
+ \partial_{--}(\eps^{0,-1\, \und a}\, G^{\,2,3}_{\und a} +
\eps^{0,1\, \und a}\, H^{\,2,1}_{\und a})\,,
\label{var2}
\eea
which depend on the same arguments as the Lagrangian in (\ref{lagr.gen2}).
A possible extra term which could be added to the r.h.s. of (\ref{var2}),
\be
\db (\eps^{0,1\, \und a}\, A^{\,2,-1}_{\und a})\,,
\ee
is reduced to the one already included, after representing
$\eps^{0,1\, \und a} = D^{0, 2}\eps^{0,-1\, \und a}$
and integrating by parts.

Using the transformation rules \p{tran3} of $\qa$  and those of $\theta$'s,
it is easy to compute the explicit form of the supersymmetry variation in
the l.h.s. of (\ref{var2}) and to find that only one constraint is actually
required for the action to be invariant:
\be
\frac{\partial{\cal L}^{2,2}_0}{\partial \qa} =
{\partial}^{0,2} {\cal L}^{1,0}_{\und a}
\label{susycon2}
\ee
where the partial harmonic derivative acts only on the explicit
harmonics $v$ in ${\cal L}^{1,0}_{\und a}$\,. A corollary of this constraint
is the following condition on ${\cal L}^{1,0\;\und a}$\,:
\be
{\partial}^{0,2}\left(\frac{\partial {\cal L}^{1,0\;\und a}}{\partial \qa}\right) = 0\,.
\label{ind1}
\ee

Another consequence of the invariance condition \p{var2} is that
the last term in (\ref{lagr.gen2}) is a total
$x$-derivative and so makes no contribution.
Indeed, the most general form of the Grassmann functions $G^{\,2,3}_{\und a}$
and $H^{\,2,1}_{\und a}$ in (\ref{var2}), compatible with the Lorentz
covariance and the fact that these functions have the dimension $-1/2$, is
\be
G^{\,2,3}_{\und a} = (\theta^{0,1})^2\, g^{1,1}\, G^{\,1,0}_{\und a}\, (\qa, u, v)\,,
\qquad
H^{\,2,1}_{\und a} = \theta^{0,1\; \und b}\, H^{\,2,0}_{\und a\; \und b}\, (\qa, u, v)\,.
\label{GH}
\ee
Substituting (\ref{GH}) into the r.h.s. of (\ref{var2}) and computing
the explicit form of $\delta\, {\cal L}^{2,2}_{(a)}$\,, we find the relations
\be
G^{\,1,0}_{\und a} = - \hat {\cal L}^{1,0}_{\und a}, \qquad
\hat {\cal L}^{1,0}_{\und a} = \eps^{\und c\, \und d}\,
\frac{\partial H^{\,2,0}_{\und d\; \und c}}{\partial q^{1,0\;\und a}}\,.
\ee
The second relation implies that the last term in (\ref{lagr.gen2})
can be expressed as $x^{--}$-derivative of the function
$H^{\,2,0}_{\und a\; \und b}$\,.
The same phenomenon occurs for the Lagrangian (\ref{lagr.gen3}), and it was
the reason why we have omitted possible terms of this kind in (\ref{lagr.gen3})
from the very beginning.

Keeping in mind this remark, let us deduce the analogous constraints on
${\cal L}^{2, 2}_{(c)}$\,. For the action \p{s3gen} (or \p{lagr.gen3})
to be invariant, the variation of this Lagrangian should be as follows
\be
\delta\,{\cal L}^{2,2}_{(c)} = \da (\eps^{-1,0\, \und i}\, A^{\,1,2}_{\und i} +
\eps^{0,-1\, \und a}\, B^{\,0,3}_{\und a})
+ \db (\eps^{-1,0\, \und i}\, C^{\,3,0}_{\und i}
+ \eps^{0,-1\, \und a}\, D^{\,2,1}_{\und a})\,.
\label{var3}
\ee
Writing the variation in the l.h.s. of \p{var3} in the explicit form,
it is straightforward to find that the component functions in
(\ref{lagr.gen3}) should obey the following system of constraints:
\bea
&&
\frac{\partial {\cal L}_0^{2,2}}{\partial \tilde{q}^{\,\und i\, \und a}}
= \partial^{2,0} {\cal L}^{0,2}_{\und i\; \und a}
+ \partial^{0,2} {\cal L}^{2,0}_{\und i\; \und a}\,, \quad
{\cal L}_{(\und i\; \und l)(\und a\; \und c)} = 0\,, \label{susycon03}\\
&&
\frac{\partial {\cal L}^{2,0}_{\und i\; \und a}}
{\partial \tilde{q}^{\,\und l\, \und c}}
- \frac{\partial {\cal L}^{2,0}_{\und l\; \und c}}
{\partial \tilde{q}^{\,\und i\, \und a}}
= \partial^{2,0} {\cal L}_{\und l\; \und c\; \und i\; \und a}\,, \quad
\frac{\partial {\cal L}^{0,2}_{\und i\; \und a}}
{\partial \tilde{q}^{\,\und l\, \und c}}
- \frac{\partial {\cal L}^{0,2}_{\und l\; \und c}}
{\partial \tilde{q}^{\,\und i\, \und a}}
= \partial^{0,2} {\cal L}_{\und i\; \und a\; \und l\; \und c}\,, \label{susycon003}\\
&& \frac{\partial {\cal L}_{\und i\; \und a\; \und k\; \und b}}
{\partial \tilde{q}^{\,\und l\, \und c}}
- \frac{\partial {\cal L}_{\und l\;\und c\; \und k\; \und b}}
{\partial \tilde{q}^{\,\und i\, \und a}}
= \frac{\partial {\cal L}_{(\und i\; \und l)[\und a\; \und c]}}
{\partial \tilde{q}^{\,\und k\, \und b}}\,, \quad
\frac{\partial {\cal L}_{\und i\; \und a\; \und k\; \und b}}
{\partial \tilde{q}^{\,\und l\, \und c}}
- \frac{\partial {\cal L}_{\und i\; \und a\; \und l\; \und c}}
{\partial \tilde{q}^{\,\und k\, \und b}}
= \frac{\partial {\cal L}_{[\und l\; \und k](\und c\; \und b)}}
{\partial \tilde{q}^{\,\und i\, \und a}}
\label{susycon333}
\eea
where we introduced the notation
\be
{\cal L}_{\und i\; \und a\; \und l\; \und c} = {\cal L}_{(\und i\; \und
l)\; (\und a\; \und c)} + {\cal L}_{(\und i\; \und l)\; [\und a\;
\und c]} + {\cal L}_{[\und i\; \und l]\; (\und a\; \und c)} +
{\cal L}_{[\und i\; \und l]\; [\und a\; \und c]}
\label{susycon0003}
\ee
(the symbols $(\;\;)$ and $[\;\;]$ mean symmetrization and antisymmetrization
with the factor $1/2$).

Below we shall identify the $\theta$-independent piece of the last term
in (\ref{susycon0003}) with the bosonic target metric of sigma model,
while the antisymmetric part of
${\cal L}_{\und i\; \und a\; \und l\; \und c}\vert_{\theta=0}$
will be identified with the torsion potential.

It is straightforward to substitute the component expansion of $\qa$\,,
eq. (\ref{bos.con2}), into (\ref{lagr.gen2}) and that of
$\tilde{q}^{\,\und i \,\und a}$\,,
eq. (\ref{bos.con4}), into (\ref{lagr.gen3}), to integrate over $\theta$'s
and harmonics with the help of the constraints (\ref{susycon2}) and
(\ref{susycon03}), (\ref{susycon003}), and to eventually obtain
the component form of the actions in $x$-space. We give here only
those parts which involve the physical bosonic and auxiliary fields.

For the action (\ref{lagr.gen2}) these pieces are as follows
\bea
&&
S^{phb}_{(a)} = \frac{1}{2}\int d^2x\;\{\,G_{i\; \und a \;j \;\und b}(q)\;
\partial_{++}q^{i\, \und a}\, \partial_{--}q^{j\, \und b} +
2\,B_{i\; \und a\; j\; \und b}(q)\; \partial_{++}q^{i\, \und a}\,
\partial_{--}q^{j\, \und b}\,\}\,, \label{comp.gen22} \\
&&
S^{auxb}_{(a)} = \frac{1}{8}\int d^2x\;G(q)\;F^a_{\und i}\,F^{\und i}_a
\label{comp.aux22}
\eea
where
\bea
&&  \label{Gq2}
G_{i\; \und a\; j\; \und b}(q) = G(q)\;\eps_{i\,j}\,
\eps_{\und a\, \und b}\,, \quad G(q) = \int du\; g(q^{1,0},u)\,, \\
&&  \label{Bq2}
B_{i\; \und a\; j\; \und b}(q) = \int du\; g(q^{1,0}, u)\;u^1_{(i}u^{-1}_{j)}\;
\eps_{\und b\, \und a}\,, \\
&& \label{g2} g(q^{1,0}, u) = \left. \frac{\partial {\cal
L}^{1,0\; \und a}}{\partial \qa} \right |_{\theta = 0}\,,
\quad q^{1,0\;\und a}\vert_{\theta = 0} = q^{i\, \und a}(x)\, u^1_i\,.
\label{gq2}
\eea
The $v^{\pm 1}_a$-independence of the function $g(q^{1,0},u)$ in \p{g2} and, hence,
of the torsion potential, follows from the constraint (\ref{ind1}).

Analogous terms for the action (\ref{lagr.gen3}) read
\bea
&&
\label{comp.gen3} S^{phb}_{(c)} =  \int d^2x\;
\{\,G_{\und i\; \und a\; \und j\; \und b} + B_{\und i\; \und a\; \und j\; \und b}\,\}\,
\partial_{++} \qc\,\partial_{--} q^{\und j\; \und b}\,, \\
&&
S^{auxb}_{(c)} = \frac{1}{4} \int d^2x\; \hat G \; F^a_i\, F^i_a
\eea
where the involved objects are the appropriate $\theta = 0$ projections
\bea
&&\label{Gq3} G_{\und i\; \und a\; \und j\; \und b} = {\cal
L}_{[\; \und i\; \und j\;]\; [\; \und a\; \und b\;]}\vert_{\theta = 0} \equiv
\eps_{\und i\, \und j}\, \eps_{\und a\, \und b}\, \hat{G}\,, \quad
\hat{G} = \frac{1}{4}\, \eps^{\und i\, \und j}\, \eps^{\und a\, \und b}\,
G_{\und i\; \und a\; \und j\; \und b}\,,  \\
&&\label{Bq3} B_{\und i\; \und a\; \und j\; \und b} = \left({\cal
L}_{[\; \und i\; \und j\;]\; (\; \und a\; \und b\;)} + {\cal
L}_{(\; \und i\; \und j\;)\; [\; \und a\; \und b\;]}\right)\vert_{\theta = 0}\,.
\eea

With the help of constraints (\ref{susycon03}), (\ref{susycon003}) one can
show that the scalar metric and torsion potential in this case are
independent of both sets of harmonic variables modulo a gauge
transformation of $B_{\und i\; \und a\; \und j\; \und b}$\,. Indeed,
the second constraint in \p{susycon03} and  constraints \p{susycon003}
together imply
$$
\partial^{2,0}{\cal L}_{[\; \und i\; \und j\;]\; [\; \und a\; \und b\;]}\vert_{\theta=0} =
\partial^{0,2}{\cal L}_{[\; \und i\; \und j\;]\; [\; \und a\; \und b\;]}\vert_{\theta=0} = 0\,,
$$
whence $\partial^{2,0}\hat{G} = \partial^{0,2}\hat{G} = 0$\,.
Further, we can rewrite the $\theta = 0$ projection of (\ref{susycon003}) as
\be
\frac{\partial {\cal L}^{2,0}_{\und i\; \und a}}{\partial q^{\und l\, \und c}}
- \frac{\partial {\cal L}^{2,0}_{\und l\; \und c}}{\partial q^{\und i\, \und a}}
= \partial^{2,0}\, B_{\und l\; \und c\; \und i\; \und a}\,, \quad
\frac{\partial {\cal L}^{0,2}_{\und i\; \und a}}{\partial q^{\und l\, \und c}}
- \frac{\partial {\cal L}^{0,2}_{\und l\; \und c}}{\partial q^{\und i\, \und a}}
= \partial^{0,2}\,B_{\und i\; \und a\; \und l\; \und c}\,.
\label{susycon0033}\\
\ee
Also, one should take into account that $B_{\und i\; \und a\; \und l\; \und c}$
is defined up to the gauge transformation
\be
\delta B_{\und i\; \und a\; \und l\; \und c} =
\frac{\partial X_{\und i\; \und a}}{\partial q^{\und l\, \und c}}
- \frac{\partial X_{\und l\; \und c} }{\partial q^{\und i\, \und a}}
\ee
where $X_{\und i\; \und a}$ is an arbitrary function of $q^{\und i \,\und a}(x)$ and
harmonics (such addition to $B$ in \p{comp.gen3} produces a total $x$-derivative).
Exploiting this gauge freedom together with the constraints \p{susycon0033}
and the $\theta = 0$ projection of the first constraint in \p{susycon03},
one can show that it is possible to choose a gauge in which
$B_{\und i\; \und a\; \und l\; \und c}$ satisfies the homogeneous harmonic constraints
\be
\partial^{2,0} \tilde B_{\und i\; \und a\; \und l\; \und c} = 0\,, \;\;\;\;
\partial^{0,2} \tilde B_{\und i\; \und a\; \und l\; \und c} = 0\,,
\label{ind3}
\ee
and so indeed does not depend on harmonics.

 The objects $G_{i\; \und a\; j\; \und b}\,,\; B_{i\; \und a\; j\; \und b}$
($ G_{\und i\; \und a\; \und j\; \und b}\,,
B_{\und i\; \und a\; \und j\; \und b}$) are, respectively, symmetric and
antisymmetric under the simultaneous
permutation of the indices
$i \leftrightarrow j, \; \und a \leftrightarrow \und b$
($\und i \leftrightarrow \und j, \; \und a \leftrightarrow \und b$)
and so they can be identified with the metric
and torsion potential on the target space.
Sometimes it is advantageous to express the second terms in (\ref{comp.gen22})
and (\ref{comp.gen3}) through the torsion field strengths which are
defined by
\be
H_{i\; \und a\; j\;  \und b\; k\; \und c} =
\frac{\partial B_{i\; \und a\; j\; \und b}}{\partial q^{k\, \und c}}
+ \frac{\partial B_{k\; \und c\; i\; \und a}}{\partial q^{j\, \und b}}
+ \frac{\partial B_{j\; \und b\; k\; \und c}}{\partial q^{i\, \und a}}
\label{h2}
\ee
and
\be
H_{\und i\; \und a\; \und j\;  \und b\; \und k\; \und c} =
\frac{\partial B_{\und i\; \und a\; \und j\; \und b}}
{\partial q^{\und k\, \und c}}
+ \frac{\partial B_{\und k\; \und c\; \und i\; \und a}}
{\partial q^{\und j\, \und b}}
+ \frac{\partial B_{\und j\; \und b\; \und k\; \und c}}
{\partial q^{\und i\, \und a}}\,.
\label{h3}
\ee
They are totally antisymmetric with respect to permutations of the quartet
pairs $i\; \und a\,,\;$ $j\;  \und b\,,\;$ $k\; \und c$ in (\ref{h2}) and
$\und i\; \und a,\;$ $\und j\;  \und b,\;$ $\und k\; \und c$ in (\ref{h3}).
For $B_{i\; \und a\; j\; \und b}$ given by eq. (\ref{Bq2})
and $B_{\und i\; \und a\; \und j\; \und b}$ given by eq. (\ref{Bq3}),
the corresponding torsion field strengths
$H_{i\; \und a\; j\;  \und b\; k\; \und c}$ and
$H_{\und i\; \und a\; \und j\;  \und b\; \und k\; \und c}$ are reduced to
\be
\label{HH2}
H_{i\; \und a\; j\;  \und b\; k\; \und c} =
\eps_{\und b\, \und c}\,
\eps_{i\,(j}\, \frac{\partial G}{\partial q^{k)\, \und a}}
+ \eps_{\und a\, \und b}\,
\eps_{k\,(i}\, \frac{\partial G}{\partial q^{j)\, \und c}}
+ \eps_{\und c\, \und a}\,
\eps_{j\,(k} \frac{\partial G}{\partial q^{i)\, \und b}}
\equiv 3 \tilde H_{i\; \und a\; j\; \und b\; k\; \und c}
\ee
where
\be
\label{H2}
\tilde H_{i\; \und a\; j\;  \und b\; k\; \und c} =
\eps_{i\,j}\, \eps_{\und a\, \und c}\,
\frac{\partial G}{\partial q^{k\, \und b}}
- \eps_{i\,k}\, \eps_{\und a\, \und b}\,
\frac{\partial G }{\partial q^{j\, \und c}}
\ee
and
\be
H_{\und i\; \und a\; \und j\; \und b\; \und k\; \und c} =
\eps_{\und b\, \und a}\, \eps_{\und k\, (\und i}\,
\frac{\partial \hat{G}}{\partial q^{\und j)\, \und c}}
+ \eps_{\und a\, \und c}\, \eps_{\und j\, (\und k}\,
\frac{\partial \hat{G}}{\partial q^{\und i)\, \und b}}
+ \eps_{\und c\, \und b}\, \eps_{\und i\, (\und j}\,
\frac{\partial \hat{G}}{\partial q^{\und k)\, \und a}}\,.\label{H33}
\ee
When deducing \p{H33}, we essentially used the constraints \p{susycon333}.

We would like to point out that in both considered cases
the geometric target space objects (metric and torsion) are expressed
through single scalar functions $G(q^{i\, \und a})$ or $\hat{G}(q^{\und i \,\und a})$
defined, respectively, by eqs. \p{Gq2} and \p{Gq3}. The only constraint
they satisfy is the four-dimensional Laplace equation
\be
\mbox{(a)}\;\;\frac{\partial^2 G}{\partial q^{i\, \und a}\, \partial q_{i\, \und a}} = 0\,,
\quad \mbox{(b)}\;\;\frac{\partial^2 \hat{G}}
{\partial q^{\und i\, \und a}\, \partial q_{\und i\, \und a}} = 0\,.
\label{Lap}
\ee
Eq. (\ref{Lap}a) follows from the definition of $G$ in \p{Gq2} and the property
$$
\frac{\partial^2}{\partial q^{i\, \und a}\, \partial q_{i\, \und a}} \sim
\frac{\partial^2}{\partial q^{1, 0\,\und a}\, \partial q_{\,\und a}{}^{-1, 0}}
- \frac{\partial^2}{\partial q^{-1, 0\,\und a}\, \partial q_{\,\und a}{}^{1, 0}}\,,
$$
which is a consequence of the completeness relation
$u^1_i u^{-1}_k - u^{-1}_i u^1_k = \varepsilon_{i\,k}$\,,
whereas eq. (\ref{Lap}b) is implied by the constraints \p{susycon333}
and the second constraint in \p{susycon03}. The same bosonic target geometry
was found in the case of the analytic twisted multiplet $q^{1,1}$ in \cite{IS}
and, in the ${\cal N}=(2,2)$ superspace formulation, in \cite{GHR}.
Thus we conclude that the most general {\it{off-shell}} ${\cal N}=(4,4)$ sigma models
associated with each twisted multiplet from the four-entry set $\hat q^{\,i\,a}\,,\,
\hat q^{\,i\, \und a}\,,\, \hat q^{\,\und i\, a}\,,\, \hat q^{\,\und i\,\und a}$ defined
in eq. \p{base} show up equivalent target geometries. In the next Subsection,
on the example of the multiplet $\hat q^{\,i\, \und a}$\,, we shall see that
the same is true for the cases when a few multiplets of the same sort are present.

Finally, as a particular case of the above general actions, we quote the
supersymmetric free actions of $\hat q^{\,1,0\; \und a}$ and $\hat{q}^{\,\und i \,\und a}$
\be
S^{free}_{(a)} \sim \int \mu^{-2,-2}\, q^{1,0}_{\,\und a} \db \qa\,, \quad
\label{free2}
\ee
\be S^{free}_{(c)} \sim \int
\mu^{-2,-2}\, \da \tilde q_{\,\und i\, \und a}\, \db \tilde q^{\,\und i\, \und a}\,.
\label{free3}
\ee

\subsection{Generalization to the case of several $\hat q^{\,1,0 \; \und a}$}

Generalizing the action of a single $q^{1,1}$ superfield to the
case of $n$ self-interacting superfields $q^{1,1\,M}$ $(M=1, \ldots , n)$ is
straightforward, see eq. (\ref{s1gen}).
Now we are going to generalize the supersymmetric action of single
$\hat q^{\,1,0 \; \und a}$ to the general case of several self-interacting
$\hat q^{\,1,0 \; \und a\, M}$\,. The supersymmeric transformation properties
of $q^{1,0 \; \und a\, M}$ are
\be
\delta\; q^{1,0 \; \und a\, M} = -\eps^{0,-1 \, \und a}\, g^{1,1\,M}\,, \quad \delta\,g^{1,1\,M} =0\,. \\
\ee
The defining harmonic constraints for each value of the index $M$ have the form
(\ref{harm.con2}).
Solving them, we find the bosonic component content of $q^{1,0 \; \und a\, M}$ as
\be
q^{1,0 \; \und a\, M}= q^{i\, \und a\,M} u^1_i + \ti \ta F^{a\,M}_{\und i} v^{-1}_a
- i(\theta^{1,0})^2 \partial_{++} q^{i\, \und a\,M} u^{-1}_i\,.
\label{Bos.con2}
\ee
Following the same line of arguments as in the construction of (\ref{s2gen}),
(\ref{lagr.gen2}), we can again take the candidate general action as an analytic
superspace integral of some function ${\cal L}^{2,2}_{(a)}$\,:
\be
S^{Gen}_{(a)} = \int \mu^{-2,-2}\, {\cal L}^{2,2}_{(a)}\,
(q^{1,0 \; \und a\, M}, g^{1,1}{}^M, \ta\,, u, v)
\label{s2Gen}
\ee
and then specify it according to the harmonic constraints as
\bea
{\cal L}^{2,2}_{(a)} &=& {\cal L}^{2,2}_0(q^{1,0 \; \und a\, M},u,v)
+ {\cal L}^{1,0\, M}_{\und a}(q^{1,0 \; \und a\, M},u,v)\, \db q^{1,0 \; \und a\, M} \nn \\
&+& \, {\cal L}^{0,-2\, MN}_{\und a\; \und b}\,(q^{1,0 \; \und a\, M},u,v)
\db q^{1,0 \; \und a\, M} \db q^{1,0\; \und b\, N}\,.
\label{lagr.Gen2}
\eea
Like in the case of single $\hat q^{\,1,0 \; \und a}$\,, we omit a possible term
with explicit $x$-derivative (and explicit $\theta$'s) in (\ref{lagr.Gen2})
because it can be shown to be a total $x$-derivative as a consequence
of requiring \p{s2Gen} to be supersymmetric.
Demanding that the variation of the Lagrangian (\ref{lagr.Gen2})
under the supersymmetry transformations is a total harmonic derivative
of an arbitrary function depending on the same arguments as
the Lagrangian itself,
\be
\delta\, {\cal L}^{2,2}_{(a)} = \db (\eps^{ 0,-1\, \und a}\, F^{2,1}_{\und a})\,,
\label{Var2}
\ee
one finds the set of constraints for the considered case:
\be
\frac{\partial{\cal L}^{2,2}_0}{\partial q^{1,0 \; \und a\, M}} = {\partial}^{0,2}
{\cal L}^{1,0\, M}_{\und a}\,,
\label{fir2}
\ee
\be
\frac{\partial {\cal L}^{1,0\, N]}_{\und b}}{\partial q^{1,0\; \und a\,[M }}
- \frac{\partial {\cal L}^{1,0\, [M}_{\und a}}{\partial q^{1,0\;\und b\, N] }}
= \partial^{0,2} {\cal L}^{0,-2\, [M N]}_{\und a\; \und b}
\label{sec2}
\ee
(one automatically gains antisymmetrization in indices $M, N$ in \p{sec2}
since the latter actually emerges multiplied by $g^{1,1\,M}g^{1,1\,N}$).
The constraint (\ref{fir2}) has the same form as in the case of single
$\hat q^{\,1,0 \; \und a}$\,. The constraint (\ref{sec2}) is new.
Let us discuss what it means.
First, from the structure of the last term in (\ref{lagr.Gen2}) one can derive
that the function ${\cal L}^{0,-2 MN}_{\und a\; \und b}$ is antisymmetric
with respect to the permutation of each pair of its indices
\be
{\cal L}^{0,-2\, MN}_{\und a\; \und b}
= - {\cal L}^{0,-2\, NM}_{\und a\; \und b}
= - {\cal L}^{0,-2\, MN}_{\und b\; \und a}
= {\cal L}^{0,-2\, NM}_{\und b\; \und a}\,.
\ee
Second, one can rewrite the l.h.s. of (\ref{sec2}) as
\be
\frac{\partial {\cal L}^{1,0\, N]}_{\und b}}{\partial q^{1,0\; \und a\, [M}}
- \frac{\partial {\cal L}^{1,0\, [M}_{\und a}}{\partial q^{1,0\; \und b\, N]}}
= \frac{1}{2}\,\frac{\partial {\cal L}^{1,0\,N] }_{(\,\und b}}
{\partial q^{1,0\; \und a)\,[M}}\,.
\ee
Thus we see that (\ref{sec2}) actually amounts to the two independent constraints
\be
\frac{\partial {\cal L}^{1,0\,[M }_{(\,\und a}}
{\partial q^{1,0\; \und b\,)\,N]}} = 0\,,\quad
\partial^{0,2} {\cal L}^{0,-2\, [M N]}_{\und a\; \und b} = 0 \;\; \Rightarrow \;\;
{\cal L}^{0,-2\, [M N]}_{\und a\; \und b} = 0\,.
\label{sec22}
\ee

Now we are prepared to write the component form of the bosonic sector of
the general action (\ref{lagr.Gen2}). After integrating over $\theta$'s
with the help of (\ref{fir2}), (\ref{sec22}), one finds
\bea
&& \kern-12mm
S^{phb}_{(a)} = \frac{1}{2} \int d^2x\; \{\,
G^{MN}_{i\; \und a\; j\; \und b}(q) \;
\partial_{++} q^{j\, \und b\,N}\, \partial_{--} q^{i\, \und a\, M}
+ 2\; B^{MN}_{i\; \und a\; j\; \und b}(q) \;
\partial_{++} q^{j\, \und b\,N}\, \partial_{--} q^{i\, \und a\, M}\,\}\,,
\label{comp.Gen2} \\
&& \kern-12mm
S^{auxb}_{(a)}(q) = \frac{1}{8} \int d^2x\; G^{MN}(q)\, F^{\,a\,M}_{\und i}\, F^{\,\und i\,N}_a
\eea
where
\bea
&&
G^{MN}_{i\; \und a\; j\; \und b}(q) =
\int du \;g^{MN}(q, u)\, \eps_{i\,j}\, \eps_{\und a\, \und b} =
\eps_{i\,j}\, \eps_{\und a\, \und b}\, G^{MN}(q)\,, \nn\\
&&
B^{MN}_{i\; \und a\; j\; \und b}(q) =
\int du\; g^{MN}(q, u)\, u^1_{(\;i} u^{-1}_{j\;)}\, \eps_{\und a\, \und b}\,, \nn\\
&&
G^{MN}(q) = \int du \; g^{MN}(q, u)\,, \quad
g^{MN}(q,u) = \left. \frac{\partial {\cal L}^{1,0\; \und a\,(M}}
{\partial q^{1,0\; \und a \,N)}}
\right |_{\theta = 0}\,.
\label{defGMN}
\eea
The objects $G^{MN}_{i\; \und a\; j\; \und b}(q)$\,,
$B^{MN}_{i\; \und a\; j\; \und b}(q)$ are, respectively, symmetric and
antisymmetric under the simultaneous permutation of the indices
$i \leftrightarrow j, \; \und{a} \leftrightarrow \und{b}, \;
M \leftrightarrow N$ and so they can be identified with the target space metric
and torsion potential. The torsion field strength is given by
\be
H^{MNT}_{i\; \und a\; j\;  \und b\; k\; \und c} =
\eps_{\und b\, \und c}\,
\eps_{i\,(j}\, \frac{\partial G^{NT}}{\partial q^{k)\, \und a\,M}}
+ \eps_{\und a\, \und b}\,
\eps_{k\,(i}\, \frac{\partial G^{MN}}{\partial q^{j)\, \und c\,T}}
+ \eps_{\und c\; \und a}\,
\eps_{j\,(k}\, \frac{\partial G^{TM}}{\partial q^{i)\, \und b\,N}}\,.
\ee

An analog of the basic functions $G(q)$, $\hat{G}(q)$ of the one-multiplet
case is the symmetric $n\times n$ matrix function $G^{MN}(q)$ through which
both the metric and torsion are expressed. From its
definition \p{defGMN} and the first constraint in \p{sec22} it is easy
to find analogs of the constraint \p{Lap} for the considered case
\be
(\mbox{a}) \;\; \frac{\partial^2 G^{MN}}{\partial q^{i\, \und a \,T}\, \partial q_{i\, \und a}{}^F} = 0\,,
\quad (\mbox{b}) \;\;\frac{\partial G^{MN}}{\partial q^{i\, \und a\,T}} -
\frac{\partial G^{TN}}{\partial q^{i\, \und a\,M}}\,= 0\,. \label{constrMN}
\ee
This is the manifestly $SU(2)\times SU(2)$ covariant form of the similar
constraints obtained in \cite{GHR} for the case of a few twisted multiplets
in the ${\cal N}=(2,2)$ superfield approach. They also coincide with the constraints
for a similar $n\times n$ metric for the case of $n$ superfields $q^{1,1\, M}$
\cite{IS}. This metric is defined  as the $\theta = 0$ projection of
$$
\tilde{G}^{MN}(q) = \int du\, dv\;
\frac{\partial^2 {\cal L}^{2,2}(q^{1,1\, T}, u, v)}{\partial q^{1, 1\,M}\, \partial q^{1,1\,N}}\,.
$$
It is straightforward to check that it satisfies the same constraints \p{constrMN},
up to the replacement $q^{i\, \und a\, T} \rightarrow q^{i\, a\,T}$\,.
Thus in both cases we are facing the same sort of the bosonic target HKT geometry.

One can construct an analogous off-shell superfield action also for several
twisted multiplet $\hat q^{\,\und i\, \und a}$, though such a construction
is somewhat more involved. The corresponding component action and bosonic
target geometry are the same as in the case of $q^{1,1\,M}$ or
$\hat q^{1,0\;\und a\, M}$ (up to the proper rearrangement of $SU(2)$
indices of the component fields).

\setcounter{equation}{0}

\section{$SU(2)\times U(1)$ WZNW sigma model of $\hat q^{\,1,0 \; \und a}$ multiplet}

In this Section we present one more explicit example of off-shell action for
$\hat q^{\,1,0 \; \und a}$ (besides the free action \p{free2}).
We shall show that the requirement of invariance under one of two ${\cal N}=(4,4)$
$SU(2)$ superconformal groups defined in Sect. 2 uniquely fixes the
$\hat q^{\,1,0 \; \und a}$ sigma model action to be that of ${\cal N}=(4,4)$
$SU(2)\times U(1)$ WZNW sigma model.

As a first step, we must find the transformation properties of $\hat q^{\,1,0 \; \und a}$
under both $SU(2)$ superconformal groups defined in Sect. 2. These transformation laws
are uniquely fixed by the requirement of preserving the harmonic constraints
(\ref{harm.con2}). Since these constraints do not respect a symmetry under the permutation
of the left and right light-cone sectors (as opposed to the $q^{1,1}$ defining constraints
\p{hc1}), the left and right components of superconformal groups have a different action
on the set $(\qa, \, g^{1,1})\,$.

The left light-cone branches of two ${\cal N}=(4,4)$ $SU(2)$ superconformal groups act
on $\qa$ and $g^{1,1}$ in the very simple manner
\bea
&& \delta_{(I)L}\, \qa = \Lambda_{(I)L}\, \qa\,,\quad
\delta_{(I)L}\, g^{1,1} = \Lambda_{(I)L}\, g^{1,1}\,, \nn \\
&& \delta_{(II)L}\, \qa = 0\,,\quad
\delta_{(II)L}\, g^{1,1} = 0
\label{con.rul1}
\eea
where the parameter $\Lambda_{(I)L}$ was defined in (\ref{sc.sol}).
The requirement of preserving the harmonic constraints (\ref{harm.con2}) under
the action of right light-cone branches of these superconformal groups
results in the following transformation laws
\bea
&& \delta_{(I)R}\, \qa = - \Lambda_{(I)}^{0,-1\; \und a}\, g^{1,1}\,, \quad
\delta_{(I)R}\, g^{1,1} = \Lambda_{(I)R}\, g^{1,1}\,,\label{con.rul2} \\
&&
\delta_{(II)R}\, \qa = - \Lambda_{(II)}^{0,-1\; \und a}\, g^{1,1}
+ q^{1,0\; \und b}\, \frac{\partial \Lambda_{(II)}^{0,1\; \und a}}{\partial \tb}
- \qa\, \frac{\partial \Lambda_{(II)}^{0,1\; \und b}}{\partial \tb}\,, \nn\\
&&
\delta_{(II)R}\, g^{1,1} = g^{1,1}\, \partial_{--} \Lambda_{(II)R} + 2i \qa\, \partial_{--}
\Lambda_{(II)\, \und a}^{0,1}\,.\label{con.rul3}
\eea
All the involved parameters $\Lambda$ were defined in Sect. 2.

Let us now specialize to a single $\hat q^{\,1,0 \; \und a}$ and construct
for it an action invariant under the superconformal groups defined by eqs.
(\ref{tran1L}) -- (\ref{iden2}), (\ref{con.rul1}) - (\ref{con.rul3}).
The free action (\ref{free2}) does not respect the full superconformal invariance,
it is invariant only under the ${\cal N}=4$ superconformal group $II$ of the left sector
and the ${\cal N}=4$ superconformal group $I$ of the right sector (the corresponding R-symmetry
$SU(2)$ groups act only on fermions and auxiliary fields in $\hat q^{\,1,0\;\und a}$).
So, if we wish to have invariance also under those ${\cal N}=4$ superconformal groups whose
$SU(2)$ subgroups affect physical bosonic fields, the corresponding invariant action
should necessarily include self-interaction.
To find its precise form, we apply the procedure which has been employed
earlier in \cite{tens} for constructing an action of improved ${\cal N}=2,$ $4D$
tensor multiplet in the harmonic analytic ${\cal N}=2$, $4D$ superspace and in \cite{IS} for
constructing a superconformal action of $q^{1,1}$ in the $SU(2)\times SU(2)$ HSS.
Let us introduce
\bea
\tilde q^{\,1,0\; \und a} = \qa - c^{1,0\; \und a}\,,\quad
c^{\pm 1,0\; \und a} = c^{i\, \und a} u^{\pm 1}_i\,, \quad
X = \tilde q^{\, 1,0\; \und a} c^{-1,0}_{\,\und a}\,,\quad
c^2 \equiv \frac{1}{2}\, c^{i\, \und a}\, c_{i\, \und a}\,. \nn
\eea
These newly defined quantities have the following inhomogeneous transformation
law under the action of the first superconformal group
\be
\delta_{(I)L}\, c^{1,0\; \und a} = \Lambda^{2,0} c^{-1,0\; \und a}\,, \quad
\delta_{(I)L}\, \tilde q^{\,1,0\; \und a} = \Lambda_{(I)L}\,
(\tilde q^{\,1,0\; \und a} + c^{1,0\; \und a}) - \Lambda^{2,0} c^{-1,0\; \und a}\,.
\label{Ionc}
\ee
Now we represent the sought superconformal action as a series in $X$
\be
S^{cs} = \frac{1}{2 \kappa^2} \int \mu^{-2,-2}\, \tilde q^{\,1,0\; \und a}\, \db\,
\tilde q^{\,1,0}_{\und a} \sum^{\infty}_{n=0} a_n X^n\,.
\label{Con.Ac1}
\ee
Using the relation
\be
\label{id}
c^{1,0}_{\und a} c^{-1,0}_{\und b} - c^{1,0}_{\und b} c^{-1,0}_{\und a}
 = \eps_{\und a\, \und b}\, c^2\,,
\ee
one can rewrite the prefactor in \p{Con.Ac1} also in terms of $X$\,:
\be
\label{id1}
\tilde q^{\,1,0\; \und a}\, \db\, \tilde q^{\,1,0}_{\und a} =
\frac{1}{c^2}\, (\da X \db X - X \da \db X)\,.
\ee
Now, keeping in mind that the newly introduced  analytic superfunction $X$
transforms inhomogeneously under the superconformal transformation,
one concludes that there is a possibility to achieve the invariance
of (\ref{Con.Ac1}) by requiring that the variations of the terms of different
order in $X$ cancel each other up to total harmonic derivatives. Namely,
we take into account the invariance of the integration measure and then
demand the homogeneous part of the variation of the second-order term
to be cancelled by the inhomogeneous part of the variation of the
third-order term, etc. Proceeding in this way, one finally proves that the action
(\ref{Con.Ac1}) is invariant provided the following recurrence relations between
the coefficients $a_n$ hold
\be
a_{n+1} = - \frac{1}{c^2} \frac{(n+2)^2}{(n+1)(n+3)} a_n\,,
\ee
whence one finds
\be
a_n = (-1)^n \left(\frac{1}{c^2}\right)^n \frac{2(n+1)}{(n+2)} a_0\,.
\ee
Introducing new variable $t\,$,
\be
t = \frac{X}{c^2}\,, \nn
\ee
it is straightforward to show that the series in (\ref{Con.Ac1}) is summed
up into the expression
\be
S_{sc} = \frac{1}{\kappa^2} \int \mu^{-2,-2}\, \tilde q^{\,1,0\; \und a}\,
\db\, \tilde q^{\,1,0}_{\und a}\,
\left\{\frac{\ln(1+t)}{t^2} - \frac{1}{t(1+t)}\right\}\,.
\label{Con.Ac2}
\ee
The Lagrangian in (\ref{Con.Ac2}) is the sought superconformally invariant
Lagrangian of the $\hat q^{\,1,0 \; \und a}$ multiplet.
Integrating by parts with the help of formulas (\ref{id}), (\ref{id1}),
one can rewrite the action (\ref{Con.Ac2}) in the more concise equivalent form
\be
S_{sc} = \frac{1}{\kappa^2} \int \mu^{-2,-2}\, \frac{1}{(1+t)^2}\,\da t\, \db t\, .
\label{Con.Ac3}
\ee
Using the transformation law (\ref{con.rul2}) of $\qa$\,, it is easy to check
the invariance of the action (\ref{Con.Ac2}) also under the right light-cone
branch of the considered superconformal group. This action is also invariant
with respect to the second of two ${\cal N}=(4,4)$ supeconformal groups defined in Sect. 2.
To demonstrate this, one should take the action in the form (\ref{Con.Ac3}) and
use the identity (\ref{id}), the constraints (\ref{harm.con2}) combined with
the constraints on $\Lambda_{(II)}^{0,1\; \und a}$\,, eqs. (\ref{sc.sol2}),
and the following commutator
\be
[\db, \frac{\partial}{\ta}] = - 2i\, \theta^{0,1}_{\und a} \partial_{--}\,.
\ee

In order to be convinced that the
action (\ref{Con.Ac2}) indeed describes ${\cal N}=(4,4)$ superextension of $SU(2)
\times U(1)$ WZNW model, we give here its component form.

Let us begin with the bosonic part of the action. It is given by a sum of
the physical and auxiliary bosonic field terms which, after integrating
over Grassmann variables, take the form
\bea
&&
S_{phb} =  \frac{1}{2\kappa^2} \int d^2x\;\{\,G_{i\; \und a\; j\; \und b}(q)\;
\partial_{++}q^{i\, \und a}\, \partial_{--}q^{j\, \und b} +
2\,B_{i\; \und a\; j\; \und b}(q)\;
\partial_{++}q^{i\, \und a}\,\partial_{--}q^{j\, \und b}\,\}\,, \nn\\
&&
S_{auxb} =  \frac{1}{8 \kappa^2} \int d^2x\;G(q)\;F^a_{\und i}\,F^{\und i}_a
\label{comp.gen2}
\eea
where
\bea
&&  \label{Gq22}
G_{i\; \und a\; j\; \und b}(q) = \int du\,dv\; g(t)\;\eps_{i\,j}\,
\eps_{\und a\, \und b}\,, \\
&&  \label{BQ2}
B_{i\; \und a\; j\; \und b}(q) = \int du\,dv\; g(t)\;u^1_{(i}u^{-1}_{j)}\;
\eps_{\und a\, \und b}\,, \\
&& \label{gq22}
G(q) = \int du\,dv\; g(t)\,, \quad
g(t) = \left. \frac{1}{(1+t)^2}\, \right |_{\theta = 0}\,.
\eea
It turns out that all the target geometry quantities present in the
Lagrangian (including its fermionic part) are eventually expressed
through the single object $G(q)\,$.
So, in order to find an explicit formula for the metric $G(q)$ we need
to calculate the harmonic integral in \p{gq22}. Following  ref. \cite{IS}
and choosing $c^2 =1$, one can fix the freedom with respect to two
independent rigid $SU(2)$ groups realized on the indices $i$ and $\und a$,
as well as with respect to two rigid $SU(2)$ groups from the left and
right branches of the superconformal group I (recall the transformation
rule \p{Ionc}), in such a way that
\be
\label{RQ}
c^{i\, \und a} = \eps^{i\, \und a}\,, \quad
q^{i\, \und a} = \eps^{i\, \und a}\, \rho (x)\,, \quad
{\rho}^2 = \frac{1}{2}\, (q^{i\, \und a}\, q_{i\, \und a})\,.
\ee
In this frame, using (\ref{id}), we find that
\be
t = (\qa - c^{1,0\; \und a})\, c^{-1,0}_{\,\und a} = (\rho - 1)\,.
\ee
Then, the calculation of the harmonic integral
yields the simple expression for the metric $G(q)$ as
\be
\label{GR}
G(q) = \int du \,dv\; g(t) = {\rho}^{-2}\,.
\ee
Parameterizing the $4 \times 4$ matrix of physical bosons as
\be
\label{par}
q^{i\, \und a} (x) =  e^{u(x)} \, \tilde q^{\,i\, \und a}(x)
\ee
where $\tilde q^{\,i\, \und a}(x)$ is an unitary $SU(2)$ matrix,
\be
\tilde q^{\,i\, \und a}\, \tilde q^{\,k} _{\,\und a} = \eps^{k\,i}\,,\quad
\tilde q^{\,i\, \und a}\, \tilde q_{\,i}^{\,\und b} = \eps^{\und b\, \und a}\,,
\ee
we find that
\be
\label{GU}
G(q) = e^{-2u(x)}\,.
\ee
So, the metric term in (\ref{comp.gen2}) is reduced to a sum of the free
Lagrangian of the field $u(x)$ and the Lagrangian of the $SU(2)$
principal chiral field sigma model
\be
G(q)\, \partial_{++} q^{i\, \und a}\, \partial_{--} q_{i\, \und a}
= 2\, \partial_{++}u\, \partial_{--}u + \partial_{++} \tilde q^{\,i\, \und a}\,
\partial_{--} \tilde q_{\,i\, \und a}\,.
\ee

In the present case, the totally antisymmetric (with respect to
the permutations of pairs of the indices $i\, \und a\,, j\, \und b\,,...$)
torsion field strength $H_{i\; \und a\; j\; \und b\; k\; \und c}$
defined by the general formula (\ref{HH2}) is given by the simple
expression
\be
H_{i\; \und a\; j\;  \und b\; k\; \und c} =
\eps_{i\,j}\, \eps_{\und a\, \und c}\,
\frac{\partial G}{\partial q^{k\, \und b}}
- \eps_{i\,k}\, \eps_{\und a\, \und b}\,
\frac{\partial G}{\partial q^{j\, \und c}}\,,
\ee
which, taking into account (\ref{GR}), is reduced to
\be
H_{i\; \und a\; j\;  \und b\; k\; \und c} = {\rho}^{-4}
(\eps_{i\,k}\, \eps_{\und a\, \und b}\,
q_{j\, \und c}
- \eps_{i\,j}\, \eps_{\und a\, \und c}\,
q_{k\, \und b})\,.
\ee
After substituting this expression into the torsion term
\be
\label{Bb}
B_{i\; \und a\; j\; \und b}\, \partial_{++} q^{i\, \und a}\, \partial_{--} q^{j\, \und b}
= \int_0^1 dt\; H_{i\; \und a\; j\; \und b\; k\; \und c}\; \partial_t\, q^{i\, \und a}\,
\partial_{++} q^{j\, \und b}\, \partial_{--} q^{k\, \und c}
\ee
and passing to the parametrization (\ref{par}), the r.h.s. of
(\ref{Bb}) takes the form
\be
\int^1_0 dt \; \partial_t \, \tilde
q_{\,i\, \und a} \, \tilde q_{\,j\, \und b} \, (\,\partial_{++} \tilde q^{\,j\, \und a}\,
\partial_{--} \tilde q^{\,i\, \und b} - \partial_{++} \tilde q^{\,i\, \und b}\,
\partial_{--} \tilde q^{\,j\, \und a}\,)\,,
\ee
which is the standard $SU(2)$ WZNW term.

Summing up the above contributions, one may write the final
expression for the bosonic part of the action (\ref{Con.Ac3}) as
\bea
S^{bos}_{sc} &=& \frac{1}{\kappa^2} \int d^2x\; \{\, (\,\partial_{++}u\, \partial_{--}u
+ \frac{1}{2}\,\partial_{++} \tilde q^{\,i\, \und a}\, \partial_{--} \tilde q_{\,i\, \und a}\,) \nn\\
&+&  \int^1_0 dt\; \partial_t \, \tilde q_{\,i\, \und a}\,  \tilde q_{\,j\, \und b}\,
(\,\partial_{++} \tilde q^{\,j\, \und a}\, \partial_{--} \tilde q^{\,i\, \und b}
- \partial_{++} \tilde q^{\,i\, \und b}\, \partial_{--} \tilde q^{\,j\, \und a}\,) \nn\\
&+& \frac{1}{8}\, e^{-2u}\, F^a_{\und k}\,F^{\und k}_a \,\}\,.
\label{boscon}
\eea

Let us now turn to the fermionic sector. The fermionic part of
the component action consists of three pieces
\be
S^{ferm}_{sc} = S_{4f} + S_{auxf} + S_{kinf}
\ee
which correspond, respectively, to the term quartic in fermionic
fields, a term involving auxiliary fields and the kinetic term
of fermions. These are as follows
\be
S_{4f} = \frac{1}{16 \kappa^2} \int d^2x\; \frac{\partial^2 G}
{\partial q^{i\, \und a}\, \partial q^{k\, \und b}}\,
\beta^i_a\, \beta^{k\,a}\, \alpha^{\und a}_{\und n}\, \alpha^{\und n\, \und b}\,,
\ee
\be
S_{auxf} =  \frac{1}{4 \kappa^2} \int d^2x\; \frac{\partial G}
{\partial q^{i\, \und a}}\, F^{\und n\, a}\,
\beta ^i_a\, \alpha_{\und n}^{\und a}\,,
\ee
\be
S_{kinf} =  \frac{1}{4i \kappa^2} \int d^2x\; \{ G\,(\alpha^{\und a}_{\und n}\,
\partial_{--} \alpha^{\und n}_{\und a}
+ \beta^i_a\, \partial_{++} \beta^a_i)
- \frac{\partial G}{\partial q^{i\, \und a}}\,
(\alpha^{\und a}_{\und n}\, \alpha^{\und n}_{\und b}\, \partial_{--} q^{i\, \und b}
+ \beta^i_a\, \beta^a_k\, \partial_{++} q^{k\, \und a})\}\,.
\ee

Using the explicit expressions (\ref{RQ}), (\ref{GR}), (\ref{GU}) for $G$,
one observes: \\

\vspace{0.2cm}
\noindent{\it (i).} After the field redefinition
\be
\tilde F^{\,\und n\, a} = F^{\und n\, a} + e^{-u}\, \tilde q^{\,k\, \und b}\,
\beta^a_k\, \alpha^{\und n}_{\und b}
\ee
the sum of $S_{4f}$ and $S_{auxf}$ is entirely cancelled by the contribution
coming from $S^{bos}_{sc}$\,. Thus the {\it off-shell} action does not contain
4-fermionic term which is present in the generic action. The full auxiliary
fields part of the action takes the simple form
\be
\label{scaux}
S^{aux}_{sc} = \frac{1}{8 \kappa^2} \int d^2x\; e^{-2u}
\tilde F_{\,\und k}^a\, \tilde F^{\,\und k}_a\,. \\
\ee

\noindent{\it (ii).} Being written through redefined fermionic fields
\be
\alpha^{i\, \und n} = e^{-u}\, \tilde q^{\,i\, \und a}\, \alpha^{\und n}_{\und a}\,,\quad
\beta^{a \und b} = e^{-u}\, \tilde q^{\,i\, \und b}\, \beta^a_i\,,
\ee
$S_{kinf}$ is reduced to a sum of the free fermionic terms
\be
\label{kinf}
S_{kinf} = \frac{i}{4\kappa^2} \int d^2x\; (\alpha^{i\, \und n}\, \partial_{--} \alpha_{i\, \und n}
+ \beta^{a\, \und b}\, \partial_{++} \beta_{a\, \und b})\,.
\ee

Up to a redefinition of $SU(2)$ indices, the full action coincides
with the component action of ${\cal N}=(4,4)$ WZNW model based on the
multiplet $q^{1,1}$ \cite{IS}.

\setcounter{equation}{0}

\section{Potential terms of $\hat q^{\,1,0 \; \und a}$ and
$\hat q^{\,\und i\, \und a}$}

Usually, the potential (or mass) terms and, in particular, mass term
in ${\cal N}=2$, $4D$ sigma models actions are generated as a result of including
central charges into the algebra of supersymmetry \cite{AGF, Gates}.
It was observed for the first time in \cite{IK} that in the ${\cal N}=(4,4)$
sigma models there exists a possibility to construct such terms without
changing the supersymmetry algebra. The explicitly elaborated example is
the ${\cal N}=4$ $SU(2)$ WZNW - Liouville system of refs. \cite{IK,IKL,GI} which is
a superconformally invariant deformation of ${\cal N}=(4,4)$ supersymmetric
$SU(2) \times U(1)$ WZNW model. As shown in \cite{IS}, in the $SU(2)\times SU(2)$
HSS the off-shell mass terms of $q^{1,1}$ multiplet are defined in a unique way
and result, at the component level, in the scalar potential fully specified by
the bosonic target metric. Here we construct similar terms for
$\hat q^{\,1,0 \; \und a}$ and $\hat q^{\,\und i\, \und a}$ multiplets and,
as an example, present a massive deformation of the superconformal action
(\ref{Con.Ac2}). Mass terms which involve twisted multiplets of different
types will be discussed in Sect. 6.

Keeping in mind that the superfunction $\qa$ and the integration measure
$\mu^{-2,-2}$ are dimensionless, the only way to construct a mass term is
to allow explicit $\theta$'s in the action. The simplest term of this
kind for the multiplet $\hat q^{\,1,0 \; \und a}$ reads
\be
S^m_{(a)} = m \int \mu^{-2,-2}\, \ti\, \ta\, C^{\,0,1}_{\und i\; \und a\; \und b}\,(u, v)\,
q^{1,0\; \und b}
\label{Smass2}
\ee
where the harmonic dependence of $C^{\,0,1}_{\und i\; \und a\; \und b}\,(u,v)$
is unspecified for the moment. It is easy to show that the requirement of
invariance of this term under the supersymmetry transformation constrains
$C^{\,0,1}_{\und i\; \und a\; \und b}$ in the following way
\be
C^{\,0,1}_{\und i\; \und a\; \und b} = \eps_{\und a\, \und b}\, C^{\,0,1}_{\und i}\,,\quad
\db C^{\,0,1}_{\und i} = 0\,,\quad \da C^{\,0,1}_{\und i} = 0\,.
\label{COnstr}
\ee
To this end, one represents the supertranslations of $\ta$ and $\ti$ as
\be
\delta \ta = \eps^{b\, \und a}\, v^1_b = \db \eps^{b\, \und a}\, v^{-1}_b\,, \quad
\delta \ti = \eps^{k\, \und i}\, u^1_k = \da \eps^{k\, \und i}\, u^{-1}_k\,,
\ee
integrates by parts with respect to $\db$, $\da$  and uses the harmonic
constraint (\ref{harm.con2}). The general solution of \p{COnstr} is
\be
C^{\,0,1}_{\und k} = C^a_{\und k}\, v^1_a
\ee
where $C^a_{\und k}$ are some constants.

Let us examine how adding of (\ref{Smass2}) to the generic action
(\ref{lagr.gen2}) affects the component structure of the latter.
After integrating over Grassmann and harmonic variables, one finds
the {\it{off-shell}} component action
\be
S^m_{(a)} = -\frac{m}{4} \int d^2x\; C^{\und k\, a}\, F_{\und k\, a}\,.
\label{mass.off}
\ee
Then, after eliminating the auxiliary fields in the sum $S_q + S^m_{(a)}$\,,
the physical component action acquires new terms which are expressed through
the scalar metric $G(q)$ defined in eq. (\ref{Gq2}).
We present here only the potential term of $q^{i\,\und a}$
\be
S^{pot}_{(a)} = \frac{m^2}{8} \int d^2x\; G^{-1}(q)\, C^{\und k\, a}\, C_{\und k\, a}\,.
\label{POta}
\ee
It is accompanied by proper Yukawa-type couplings with fermionic fields.
Note that \p{POta} yields a nontrivial scalar potential for bosonic fields
(including a possible mass term) only for non-constant function $G(q)$\,;
so no mass terms can be generated in this way starting from the free kinetic
action of single twisted multiplet. Yet, such terms can be generated in
the system of two twisted multiplets of different types, see Sect. 6.

We wish to point out that the off-shell term \p{Smass2}, being simplest,
is at the same time unique. Allowing any higher powers of $q^{1,0\;\und a}$\,,
and/or of analytic Grassmann coordinates, would require that harmonic functions
with negative $U(1)$ charges must be included, and the harmonic differential
constraints imposed on these functions by supersymmetry can be shown
to make them vanish.

The mass term for $\hat q^{\,\und i\, \und a}$ multiplet can be
written in the following form
\be
S^m_{(c)} = m \int \mu^{-2,-2}\,\ti\, \ta\,
C^{\,1,1}_{\und i\; \und a\; \und k\; \und b}\,(u,v)\, \tilde q^{\,\und k\, \und b}\,,
\label{Smass1}
\ee
with $C$ being  a set of harmonic-dependent constants which are arbitrary
for the moment. As in the case of $\hat q^{\,1,0 \; \und a}$\,,
it is easy to show that the supersymmetry condition
\be
\delta\, S^m_{(c)} = 0
\ee
is satisfied provided the harmonic functions $C$
have the structure
\be
C^{\,1,1}_{\und i\; \und a\; \und k\; \und b}
= C^{\,1,1} \eps_{\und i\, \und k}\, \eps_{\und a\, \und b}
\ee
and obey the harmonic constraints
\be
\da C^{\,1,1} = 0\,, \quad  \db C^{\,1,1} =0\,,
\ee
which have the simple unique solution $C^{\,1,1} = C^{i\,a}\, u^1_i\, v^1_a$\,.
After integrating over Grassmann and harmonic variables the corresponding
{\it off-shell} component action reads
\be
S^m_{(c)} = -\frac{m}{4} \int d^2x\; C^{k\,a}\, F_{k\,a}\,.
\ee
The potential term of $\hat q^{\,\und i\, \und a}$\,, which arises in
the sum of $S_q+S^m_{(c)}$ after eliminating the auxiliary fields,
is expressed through the metric $\hat{G}(q)$ (\ref{Gq3})
\be
S^{pot}_{(c)} = \frac{m^2}{16} \int d^2x\; \hat{G}^{-1}\, C^{k\,a}\, C_{k\,a}\,.
\ee

We observe that the potential (mass) terms of new types of twisted multiplet
have the same form as that given in \cite{IS} for the case of the $q^{1,1}$
multiplet (up to the proper replacements of $SU(2)$ indices).

As the last example of this Section, we discuss a massive deformation
of the superconformal action (\ref{Con.Ac3}):
\be
\label{smc}
S^m_{sc} =  \frac{1}{\kappa^2} \int \mu^{-2,-2}\,
\left\{ \frac{1}{(1+t)^2}\,\da t\, \db t  + 2\,m\, \ti\, \ta\, C^{\,0,1}_{\und i}\,
q^{1,0}_{\und a} \right\}\,.
\ee
In terms of component fields the mass term in (\ref{smc}) is
\be
S_m = - \frac{m}{2} \int d^2x\;  F^{\und i\, a}\, \eps_{\und i\, a}\,.
\ee
After eliminating auxiliary fields, it gives rise to the following physical component
action of deformed $SU(2) \times U(1)$ WZNW sigma model
\be
S_{sc(m)} = S^{bos}_{sc}(F_{\und i\, a} = 0) + S_{kinf} + S_m
\ee
where $S^{bos}_{sc}$ and $S_{kinf}$ are given by eqs. (\ref{scaux}), (\ref{kinf}) and
\be
S_{m} = \frac{1}{\kappa^2} \int d^2x\; \left\{ m^2\, e^{2u} +
\frac{m}{2}\, e^{-u}\, \eps^{\und k\, b}\,
\tilde q_{i\, \und a}\, \beta^i_b\, \alpha_{\und k}^{\und a} \right\}\,.
\ee
After rescaling the fields as
\be
\alpha \rightarrow 2 \alpha\,, \quad \beta \rightarrow 2 \beta\,,\quad F \rightarrow 2 F\,,
\ee
one finds that the resulting piece in the full action coincides, up to an overall normalization,
with the analogous one for the superconformal WZNW model of $q^{1,1}$ \cite{IS}.
The superconformal properties of the modified action are also the same as in \cite{IS},
up to the proper reshuffling of superconformal groups in the left and right sectors.

\setcounter{equation}{0}

\section{The actions with few types of twisted multiplets}

This Section is devoted to the proof that the general sigma model action
of a pair of two different twisted
multiplets is split into a sum of two actions, each involving only one multiplet.
From this result (and its extension to the case of larger number of multiplets)
we conclude that it is impossible to construct a nontrivial
supersymmetric sigma model action which would contain interactions among
different twisted multiplets. Nevertheless, it turns out possible to construct
mass terms including  the pairs of multiplets which are ``dual'' to each other
in the sense defined in Sect. 3. We show that the structure of these terms
is uniquely fixed by supersymmetry, as in the case of mass terms
for separate multiplets discussed in the previous Section.

\subsection{Sigma model actions}

To proceed, let us again apply to the description of our four
twisted supermultiplets in the original ${\cal N}=(4,4)$, $2D$ superspace.
As given in \p{base}, each of these multiplets carries
two indices of the full $SO(4)_L \times SO(4)_R$ automorphism
group of the theory.  One of these indices, the left index $i$ or $\und i$,
corresponds to one of two $SU(2)$ factors of $SO(4)_L$, while another,
right index $a$ or $\und a$, corresponds to one of two $SU(2)$ factors of
$SO(4)_R$. Since $SU(2)$ groups in a given light-cone sector are on completely
equal footing, and the left and right sectors are related to each other via
the reflection $+ \leftrightarrow -$, there exist only {\it two} non-equivalent
options of singling out a pair in the set \p{base}. One of these possibilities
is to pair multiplets having one $SU(2)$ index in common, e.g. $\hat q^{\,i\,a}$
and $\hat q^{\,i\, \und a}$\,, $\hat q^{\,i\,a}$ and $\hat q^{\,\und i\, a}$\,, etc.
Another possibility is to pair those having no $SU(2)$ indices in common at all.
In this case there are only two variants: $\hat q^{\,i\,a}$ and
$\hat q^{\,\und i\,\und a}$\,, or $\hat q^{\,i\, \und a}$ and $\hat q^{\,\und i\, a}$\,.

Keeping in mind these equivalences between various choices, we, without loss
of generality, can restrict our attention to considering most general actions
of the pairs $(\hat q^{\,i\,a}\,, \;\hat q^{\,i\, \und a})$ and
$(\hat q^{\,i\, \und a}\,,\; \hat q^{\,\und i\, a})$
as the two essentially different possibilities.
In the $SU(2)\times SU(2)$ analytic HSS the most general candidate Lagrangians
for these two options are given by
\bea
&& (I)\;\;\;{\cal L}^{2,2}_I \left(q^{1,1}\,,\; q^{1, 0\;\und a}\,, \; g^{1,1}\,,\;
\theta^{0,1\; \und a}\,, u^{\pm 1}_i, v^{\pm 1}_b\right)\,, \label{2I} \\
&& (II)\;\;\;{\cal L}^{2,2}_{II} \left(q^{0,1\;\und i}\,,\; q^{1, 0\;\und a}\,, \; f^{1,1}\,,\;
g^{1,1}\,, \;\theta^{1,0\; \und i}\,, \theta^{0,1\; \und a}\,, u^{\pm 1}_i, v^{\pm 1}_b\right)\,.
\label{2II}
\eea
This choice of two non-equivalent pairs is optional. We prefer it just because
it is technically easier compared to other possible choices.

Before turning to the general case, let us start with instructive simple
examples of actions which are bilinear in the twisted multiplet superfields.
There are two essentially different types of such actions: the actions
containing only one kind of multiplet and those containing two different
kinds.

As an example of the first type of actions, we consider the most general
quadratic action of the $\hat q^{\,1,0 \; \und a}$ multiplet. Taking into
account the harmonic constraints (\ref{harm.con2}) and the freedom of
integrating by parts with respect to harmonic derivatives, it can be
written in the analytic HSS as
\be
S^{quad} = \int \mu^{-2,-2}\,\{\, C^{\,0,2}_{(\und a\; \und b)}(v)\, q^{1,0\;\und a}\, q^{1,0\;\und b}
+ C^{\,0,0}_{[\und a\; \und b]}(v)\, q^{1,0\; \und b} \db \qa\, \}\,.
\ee
Requirement of ${\cal N}=(4,4)$ supersymmetry amounts to the following conditions
on the harmonic-dependent functions $C(v)$
\be
C^{\,0,2}_{(\und a\; \und b)} = 0\,,\quad
\db C^{\,0,0}_{[\und a\; \und b]} = 0\,,\quad
\Rightarrow \quad C^{\,0,0}_{[\und a\; \und b]} \sim \eps_{\und a\; \und b}\,\nn
\ee
whence the free action \p{free2} is unambiguously recovered.
The general actions quadratic in other sorts of multiplets from the set
\p{base} are also reduced to the relevant free actions.

As an example of the second possibility, we consider a bilinear  action of
the pair $(q^{1,1}, \qa)$
\be
\tilde S^{quad} = \int \mu^{-2,-2}\,\{\, C^{\,0,1}_{\und a} q^{1,1} \qa
+ B^{\,0,-1}_{\und a} q^{1,1} \db \qa\, \}
\label{II2}
\ee
where $C$ and $B$ are arbitrary harmonic constants. Keeping in mind
the defining constraint \p{hc1} for $q^{1,1}$\,, the second term is reduced
to the first one modulo a total harmonic derivative.
Then the requirement of ${\cal N}=(4,4)$ supersymmetry leads to
\be
C^{\,0,1}_{\und a} = 0\,, \nn
\ee
i.e. the supersymmetry requires \p{II2} to vanish.

Inspecting the bilinear actions of other pairs of different multiplets,
it is easy to prove that the requirement of invariance under the ${\cal N}=(4,4)$
supersymmetry transformations also implies these actions to vanish.

\newpage

\noindent{\it Action for $\hat q^{\,i\,a}$ and $\hat q^{\,i\, \und a}$\,}. \\

We start with examining the first option in \p{2I}, \p{2II}.
As in the case of one multiplet $\hat q^{\,1,0\;\und a}$ (Sect. 3),
the use of constraints \p{harm.con2} leads us to the following most general
form of the action with ${\cal L}^{2,2}_I$ given in \p{2I}
\bea
S = \int \mu^{-2,-2} \left\{ \,{\cal L}^{2,2}_0(q^{1,1}, \qa,u,v) +
{\cal L}^{1,0}_{\und a}(q^{1,1}, \qa,u,v)\, \db \qa \,\right\}.
\label{1L}
\eea
The superfield $q^{1,1}$ enters the superpotentials in \p{1L}
as an extra argument, and this dependence cannot
affect the process of deriving the constraints imposed on
these potentials by supersymmetry since $q^{1,1}$ is
a scalar superfield. Thus, the requirement that variation of the action
(\ref{1L}) under the supersymmetry transformations is proportional
to a total harmonic derivative gives rise to a single constraint
which coincides with (\ref{susycon2}).

To reveal all consequences of this constraint in the case under
consideration, we should plug into the component structure of the
action \p{1L}. We limit our consideration to the bosonic part; the
conclusions we shall arrive at are equally valid for the fermionic
part.

After integrating over the Grassmann coordinates, we find that the
terms with $x$-derivatives contain both diagonal and
non-diagonal pieces
\bea
S_{kin}^{bos} &=&  \int d^2x\; du\,dv\; \left\{\, -
\frac{\partial^2 {\cal L}^{2,2}_0}{(\partial q^{1,1})^2}\, \partial_{++} q^{i\,a}\,
\partial_{--} q^{j\,b}\, u^{-1}_i u^1_j v^1_a v^{-1}_b
+ \frac{\partial {\cal L}^{1,0}_{\und a}}{\partial q^{1,0 \und b}}\,
\partial_{++} q^{j\, \und b}\, \partial_{--} q^{i\, \und a}\, u^1_i u^{-1}_j\right. \nn \\
&-& \frac{\partial^2 {\cal L}^{2,2}_0}{\partial q^{1,1} \partial \qa}\, \partial_{++} q^{i\, \und a}\,
\partial_{--} q^{j\,b}\, u^1_j u^{-1}_i v^{-1}_b +
\frac{\partial {\cal L}^{1,0}_{\und a}}{\partial q^{1,1}}\,
\partial_{++} q^{j\,b}\, \partial_{--} q^{i\, \und a}\, u^1_i u^{-1}_j v^1_b \nn\\
&-&\left. \frac{\partial {\cal L}^{2,2}_0}{\partial q^{1,1}}\,
\partial_{++} \partial_{--} q^{i\,a}\, u^{-1}_i v^{-1}_a +
{\cal L}^{1,0}_{\und a}\, \partial_{++} \partial_{--} q^{i\, \und a}\,
u^{-1}_i \,\right\}. \label{555}
\eea
Integrating by parts in the last
two terms in \p{555} and using the constraint (\ref{susycon2}),
one can easily check that the kinetic bosonic part of the action
takes a diagonal form, i.e. reduces to a sum of kinetic terms of
the physical bosons from the multiplets $q^{1,1}$ and $\hat q^{\,1,0\; \und a}$\,.
To obtain the auxiliary field part of the bosonic component
action, we integrate over Grassmann coordinates with the help of
the constraint (\ref{susycon2}) and find that this part is also
reduced to a sum of two diagonal pieces. Each of these pieces
coincides with the auxiliary field part of the bosonic component
action of the relevant multiplet. Thus, collecting the results of
our calculations for the kinetic and the auxiliary pieces of the
action, we conclude that the bosonic component action for the pair
of multiplets splits into a sum of two parts
\bea
&&
\kern-12mm S_1 = \frac{1}{2} \int d^2x\; \left \{\,G_{i\;a\;j\;b}\,
\partial_{++} q^{i\,a}\,\partial_{--} q^{j\,b}
+ 2\, B_{i\;a\;j\;b}\, \partial_{++} q^{i\,a}\,\partial_{--} q^{j\,b}
+ \frac{1}{4}\, G_1\, F^{\und k}_{\und b}\, F_{\und k}^{\und b} \,\right \}, \label{NN1}\\
&&
\kern-12mm S_2 = \frac{1}{2} \int d^2x\; \left \{\,G_{i\; \und a\; j\; \und b}\,
\partial_{++} q^{i\, \und a}\, \partial_{--} q^{j\, \und b}
+ 2\, B_{i\; \und a\; j\; \und b}\, \partial_{++}\, q^{i\, \und a} \partial_{--} q^{j\, \und b}
+ \frac{1}{4}\, G_2\, F^{\und i}_{a}\, F_{\und i}^{a} \,\right \} \label{NN2}
\eea
where
\bea
&&
\kern-10mm
G_{i\;a\;j\;b} =
\int du\, dv\, g_1(q,u,v)\, \eps_{i\,j}\, \eps_{a\,b}\,, \quad
B_{i\;a\;j\;b} = \int du\, dv\, g_1(q, u, v)\, \eps_{b\,a}\, u^1_{(i} u^{-1}_{j)}\,, \label{Bnew}\\
&&
\kern-10mm
G_{i\; \und a\; j\; \und b} = \int du\, dv\, g_2(q,u,v)\, \eps_{i\,j}\, \eps_{\und a\, \und b}\,, \quad
B_{i\; \und a\; j\; \und b} = \int du\, dv\, g_2(q,u,v)\, \eps_{\und b\, \und a}\, u^1_{(i} u^{-1}_{j)}\,,
\label{Bnew1}\\
&&
\kern-10mm
G_1 = \int du\, dv\, g_1\,, \quad
g_1(q,u,v) = \left. \frac{\partial^2 {\cal L}^{2,2}_0 (q^{1,1}, \qa, u, v)}
{\partial q^{1,1}\, \partial q^{1,1}} \right|_{\theta = 0}\,, \nn\\
&&
\kern-10mm
G_2 = \int du\, dv\, g_2\,,
\quad g_2(q,u,v) = \left.
\frac{\partial {\cal L}^{1,0\; \und a}(q^{1,1}, \qa, u)}{\partial \qa} \right|_{\theta = 0}
\nn
\eea
(see \p{567} for the definition of bosonic components of $q^{1,1}$).

Let us make a remark about the structure of the torsion
potential. The expression for $B_{i\,a\,j\,b}$ given by eq. \p{Bnew} differs from the
one originally obtained in \cite{IS} (we denote it $\tilde B_{i\,a\,j\,b}$).
On the other hand, the torsion potential is defined up to the gauge
transformation
\be
B^{\,\prime}_{i\,a\,j\,b} =  B_{i\,a\,j\,b} + \frac{\partial X_{i\,a}}{\partial q^{j\,b}} -
\frac{\partial X_{j\,b}}{\partial q^{i\,a}}\,,
\ee
which leaves invariant the torsion field strength. It is easy to see
that the difference between $B_{i\,a\,j\,b}$ defined in (\ref{Bnew}) and
$\tilde B_{i\,a\,j\,b}$ of ref. \cite{IS} is just the above gauge
transformation corresponding to the special choice of gauge parameter
\be
X_{i\,a} = \int du\, dv \,\frac{\partial {\cal L}^{2,2}}{\partial
q^{1,1}}\, u^1_i v^1_a\,. \nn
\ee

Keeping in mind this remark,  we conclude  that the actions $S_1$ and $S_2$
given by eqs. (\ref{NN1}) and (\ref{NN2}) have the same structure
as the sigma model actions of separate multiplets. But, unlike the
previous case, the scalar functions $G_1$ and $G_2$\,, as well as
the torsion potentials, can in principle bear dependence on physical
bosonic fields of both multiplets. Nevertheless, it is easy
to check that both scalar functions and torsion potentials are
independent of $q^{ia}$ and $q^{i \und a}$\,, for $S_2$
and $S_1$\,, respectively.

Indeed, using the constraint (\ref{susycon2}) at $\theta =0$
and integrating by parts in harmonic integral, one finds
\bea
\frac{\partial G_1}{\partial q^{i\, \und a}}  &=&
\int du\, dv \, \frac{\partial^3 {\cal L}^{2,2}_0}{(\partial q^{1,1})^2\, \qa}\, u^1_i =
\int du\, dv \,\partial^{0,2}\, \frac{\partial^2 {\cal L}^{1,0}_{\,\und a}}
{(\partial q^{1,1})^2}\, u^1_i = 0\,, \nn\\
\frac{\partial B_{i\,a\,j\,b}}{\partial q^{k\, \und d}} &=& \int du\, dv \,
\frac{\partial^3 {\cal L}^{2,2}_0}{(\partial q^{1,1})^2\, q^{1,0 \;
\und d}}\, \eps_{b\,a}\, u^1_k u^1_{(i} u^{-1}_{j)} \nn\\
&=& \int du\, dv \,
\partial^{0,2}\, \frac{\partial^2 {\cal L}^{1,0}_{\,\und
d}}{(\partial q^{1,1})^2}\, \eps_{b\,a}\, u^1_k u^1_{(i} u^{-1}_{j)} = 0\,.
\label{XX}
\eea

The same conclusions about splitting into a sum of separate actions can be made
for the terms including fermionic fields. Thus the superfield action (\ref{1L})
actually amounts to a sum of superfield actions for $q^{1,1}$
and $\hat q^{\,1,0\;\und a}$\,.

\vspace{0.5cm}

\noindent{\it Action for $\hat q^{\,i\, \und a}$ and $\hat q^{\,\und i\, a}\,$.}\\

In the general case the initial action of the multiplets $\hat q^{\,i\, \und a}$
and $\hat q^{\,\und i\, a}$ can be written in the analytic subspace (\ref{AS})
in the form
\be \label{sd} S = \int \mu^{-2,-2}\, {\cal L}^{2,2}_{II}\,(\qb,
\qa, f^{1,1}, g^{1,1}, \ti, \ta, u, v )\,.
\ee
As discussed in the Subsection 3.2, using the constraints (\ref{harm.con2}),
(\ref{harm.con3}), we can unfold the Largangian in (\ref{sd}) in such
a way that it involves only the superfunctions $\qa$ and $\qb$
\bea
{\cal L}^{2,2} &=& {\cal L}^{2,2}_0(\qa, \qb, u, v)
+ {\cal L}^{-1,-1}_{\und i\; \und a}(\qa, \qb, u, v)\, \da \qb\, \db \qa \nn\\
&+& {\cal L}^{1,0}_{\,\und a}(\qa, \qb, u, v)\, \db \qa
+ {\cal L}^{0,1}_{\,\und i}(\qa, \qb, u, v)\, \da \qb \,.
\label{SD}
\eea
To find the constriants on the superpotentials in (\ref{SD}),
we once again demand that the variation of (\ref{SD}) under the supersymmetry
transformation is a sum of total harmonic derivatives of arbitrary
functions with the proper harmonic $U(1)$ charges:
\be
\delta\, {\cal L}^{2,2} = \da\, (\eps^{-1,0\, \und i}\, A^{\,1,2}_{\und i}
+ \eps^{0,-1\, \und a}\, B^{\,0,3}_{\und a})
+ \db\, (\eps^{-1,0\, \und i}\, C^{\,3,0}_{\und i}
+ \eps^{0,-1\, \und a}\, D^{\,2,1}_{\und a})\,.
\label{vdoubl}
\ee
The functions in the r.h.s. of (\ref{vdoubl}) depend on the same arguments
as the Lagrangian in (\ref{sd}).
Comparing the coefficients of $\eps^{1,0\, \und i}$ and $\eps^{0,1\, \und a}$
in both sides of (\ref{vdoubl}),
we find that the functions $A$ and $D$ have the following structure
\be
A^{\,1,2}_{\und i} = - f^{1,1}\, ({\cal L}^{0,1}_{\,\und i} +
{\cal L}^{-1,-1}_{\und i\; \und a} \da \qb)\,, \quad
D^{\,2,1}_{\und a} = -g^{1,1}\, ({\cal L}^{1,0}_{\,\und a} +
{\cal L}^{-1,-1}_{\und i\; \und a} \db \qa)\,.
\ee
Then, it is easy to demonstrate that the most general structure of the Grassmann
functions $B$ and $C$ in (\ref{vdoubl}), compatible with the Lorentz covariance
and dimensional considerations, is given by
\be
B^{\,0,3}_{\und a} = g^{1,1}\, (b^{\,-1,2}_{\und a} + b^{\,-3,1}_{\und i\; \und a} \da \qb)\,, \quad
C^{\,3,0}_{\und i} = f^{1,1}\, (c^{\,2,-1}_{\und i} + c^{\,1,-3}_{\und i\; \und a} \db \qa)
\label{BC}
\ee
where the functions $b$'s and $c$'s depend on $\qa$\,, $\qb$ and harmonics $u, v$.
The set of constraints which follows from (\ref{vdoubl}) contains these functions
\bea
\label{cd}
&&
\frac{\partial {\cal L}_0^{2,2}}{\partial \qa}
= \partial^{0,2} {\cal L}^{1,0}_{\und a} - \partial^{2,0} b^{\,-1,2}_{\und a}\,, \quad
\frac{\partial {\cal L}_0^{2,2}}{\partial \qb}
= \partial^{2,0} {\cal L}^{0,1}_{\und i} - \partial^{0,2} c^{\,2,-1}_{\und i}\,, \nn\\
&&
\frac{\partial {\cal L}^{0,1}_{\und i}}{\partial \qa}
=  \partial^{0,2} {\cal L}^{-1,-1}_{\und i\; \und a}
- \frac{\partial b^{\,-1,2}_{\und a}}{\partial \qb}
- \partial^{2,0} b^{\,-3,1}_{\und i\; \und a}\,, \nn\\
 &&
\frac{\partial {\cal L}^{1,0}_{\und a}}{\partial \qb}
= \partial^{2,0} {\cal L}^{-1,-1}_{\und i\; \und a}
- \frac{\partial c^{\,2,-1}_{\und i}}{\partial \qa}
- \partial^{0,2} c^{\,1,-3}_{\und i\; \und a}\,.
\eea
The presence of these additional harmonic functions is the difference
between this system of constraints and the unique constraint (\ref{susycon2})
for the case of the previous pair of multiplets.
However, a closer inspection of the harmonic charge structure of
functions $b$ and $c$ leads to the conclusion that these functions
can be eliminated  from the constraints \p{cd} after a proper redefinition
of the superfield potentials ${\cal L}$'s. Thus, without loss of generality,
one can set all these functions equal to zero in (\ref{cd})
\bea
&&
\frac{\partial {\cal L}_0^{2,2}}{\partial \qa}
= \partial^{0,2} {\cal L}^{1,0}_{\,\und a}\,, \quad
\frac{\partial {\cal L}^{1,0}_{\,\und a}}{\partial \qb}
= \partial^{2,0} {\cal L}^{-1,-1}_{\und i\; \und a}\,, \nn\\
&&
\frac{\partial {\cal L}_0^{2,2}}{\partial \qb}
= \partial^{2,0} {\cal L}^{0,1}_{\,\und i}\,, \quad
\frac{\partial {\cal L}^{0,1}_{\,\und i}}{\partial \qa}
=  \partial^{0,2} {\cal L}^{-1,-1}_{\und i\; \und a}\,.
\label{cdoubl}
\eea

As in the previous case, we now need to descend to the components.
The parts of the component action involving the physical and auxiliary bosonic fields
are obtained in the standard way. After substituting the component expansion of
$\qa$\,, eq. (\ref{bos.con2}), and $\qb$\,, eq. (\ref{bos.con3}),
into (\ref{SD}) and integrating over $\theta$'s, one finds that:\\

\noindent{\it (i).} The auxiliary boson part of the action consists of the two pieces
\be
\label{AUX}
S_{auxb}^1 = \frac{1}{8} \int d^2x\;G(q^{i\,\und a}, q^{\und i\, a})\;F^a_{\und i}\,F^{\und i}_a\,, \quad
S_{auxb}^2 = \frac{1}{8} \int d^2x\;\tilde G(q^{i\,\und a}, q^{\und i\, a})\;F^k_{\und b}\,F^{\und b}_k
\ee
where the scalar functions $G(q)$, $\tilde G(q)$
are defined by
\be
\label{A1}
G(q^{i\,\und a}, q^{\und i\, a}) = \int du\, dv \;g(q, \tilde q, u)\,, \quad
g(q, \tilde q, u) = \left. \frac{\partial {\cal L}^{1,0\; \und a}}
{\partial \qa} \right |_{\theta = 0}\,,
\ee
\be
\label{A2}
\tilde G(q^{i\, \und a}, q^{\und i\, a}) = \int du\,dv \;\tilde g(q, \tilde q, v)\,, \quad
\tilde g(q, \tilde q, v) = \left. \frac{\partial {\cal L}^{0,1\; \und k}}
{\partial q^{0,1\; \und k}} \right |_{\theta = 0}\,.
\ee
Here we denoted
\be
q \sim \left. \qa
\right |_{\theta = 0}\,, \quad
\tilde q \sim \left.  q^{0,1\; \und k}
\right |_{\theta = 0}\,.
\ee

The following remark is to the point here. From a straightforward calculation
making use of the constraints (\ref{cdoubl}), one finds that the off-diagonal
bilinear terms of the auxiliary fields coming from the multiplets
$\hat q^{\,1,0\;\und a}$ and $\hat q^{\,0,1\; \und i}$ are cancelled among
themselves. Thus, the total action for the auxiliary bosons part is reduced
to a sum of two terms
\be
S_{auxb} = S_{auxb}^1 + S_{auxb}^2\,.
\ee
The form of each term is the same as in the case of the corresponding single multiplet.
As a difference, the functions $G(q, \tilde q)$, $\tilde G(q, \tilde q)$ can now
depend on two different sets of physical bosonic fields, those from $\qa$ and $\qb$\,.
We shall return to this issue later.\\

\noindent{\it (ii).} After integrating over $\theta$'s, the physical boson part
takes the form
\bea
\label{Phys}
S_{phb}
&=&  \int d^2x\; du\,dv \, \left \{\,
{\cal L}^{1,0}_{\,\und a}\, \partial_{++} \partial_{--} q^{i\, \und a}\, u^{-1}_i
+ \frac{\partial {\cal L}^{1,0}_{\,\und a}}{\partial q^{1,0\; \und b}}\,
\partial_{++} q^{k\, \und b}\, \partial_{--} q^{i\, \und a}\, u^1_i u^{-1}_k \right.\nn\\
&+&
{\cal L}^{0,1}_{\,\und i}\, \partial_{++} \partial_{--} q^{\und i\, a}\, v^{-1}_a
+ \frac{\partial {\cal L}^{0,1}_{\,\und k}}{\partial q^{0,1 \und\; i}}\,
\partial_{++} q^{\und k\, b}\, \partial_{--} q^{\und i \,a}\, v^1_b v^{-1}_a \nn\\
&-& \left.
\frac{\partial^2 {\cal L}^{2,2}_0} {\partial \qa \; \partial q^{0,1\; \und k}}
\,\partial_{++} q^{i\, \und a}
\,\partial_{--} q^{\und k\, b}\, u^{-1}_i v^{-1}_b
- {\cal L}^{-1,-1}_{\und k\; \und a}\, \partial_{++} q^{\und k\, b}\,
\partial_{--} q^{i\, \und a}\, u^1_i v^1_b \, \right \}.
\eea
As in the previous case, \`a priori it includes mixed terms with $x$-derivatives.

Nevertheless, the constraints (\ref{cdoubl}) can be used to show
that the off-diagonal terms do not contribute. Thus \p{Phys} is diagonalized
\be
S_{phb} = S_{phb}^1 + S_{phb}^2
\ee
where
\be
S_{phb}^1 = \frac{1}{2} \int d^2x\;\{\,G_{i\; \und a\; j\; \und b}\,(q^{i\,\und a}, q^{\und i\, a})\;
+ 2\,B_{i\; \und a\; j\; \und b}\,(q^{i\,\und a}, q^{\und i\, a}) \,\}\,
\partial_{++}q^{i\, \und a}
\,\partial_{--}q^{j\, \und b}\,,
\ee
\be
S_{phb}^2 = \frac{1}{2} \int d^2x\;\{ \,\tilde G_{\und i\; a\; \und j\; b}\,(q^{i\,\und a}, q^{\und i\, a})\;
+ 2\,\tilde B_{\und i\; a\; \und j\; b}\,(q^{i\,\und a}, q^{\und i\, a}) \,\}\,
\partial_{++}q^{\und i\, a}\,
\partial_{--}q^{\und j\, b}
\ee
and
\bea
&&
G_{i\; \und a\; j\; \und b}(q) = \eps_{i\,j}\, \eps_{\und a\, \und b}\,G(q)\,, \;\;
B_{i\; \und a\; j\; \und b}\,(q)
= \int du\,dv \; \eps_{\und b\, \und a}\, u^1_{(i} u^{-1}_{j)}\, g(q, \tilde q, u)\,, \\
&&
\tilde G_{\und\; i\; a\; \und j b}\,(q)
= \eps_{\und i\, \und j}\, \eps_{a\, b}\, \tilde G(q)\,,\;\;
\tilde B_{\und i\; a\; \und j\; b}\,(q)
= \int du\,dv \; \eps_{\und i\, \und j}\, v^1_{(a} v^{-1}_{b)}\, \tilde g(q, \tilde q, v)\,.
\eea

We see that at the component level the sigma model action of the pair of multiplets
(\ref{SD}) is also reduced to a sum of two independent sigma model actions.
A difference of these actions from the original sigma model actions for each
multiplet is the presence of mixed dependence on both sets of physical bosons
$q^{i\, \und a}$ and $q^{\und i\, a}$ in each metric function $G$ and $\tilde G$
and in the torsion potentials. Let us now demonstrate that the supersymmetry
constraints actually require both the metric functions and torsion potentials
to depend only on their ``own'' types of the physical bosons.

We show this for the function $G$ and the corresponding torsion term $B$\,.
Let us consider the derivative
\be
\frac{\partial G(q)}{\partial q^{\und i\, a}}
= \int du\, dv\,  \frac{\partial^2 {\cal L}^{1,0\; \und a}}
{\partial \qa\, \partial \qb}\, v^1_a\,.\label{DirI}
\ee
Using the constraint
\be
\frac{\partial {\cal L}^{1,0}_{\,\und a}}{\partial \qb}
= \partial^{2,0} {\cal L}^{-1,-1}_{\und i\; \und a}\,, \nn
\ee
we find that, after integrating by parts with respect to the
harmonic derivative $\partial^{2,0}$, the r.h.s. of \p{DirI}
vanishes, hence $G(q)$ does not depend on $q^{\und i a}$\,.

Analogously, for the torsion potential $B_{i\, \und a\, j\, \und b}$
one finds
\be
\frac{\partial}{\partial q^{\und k\, c}} B_{i\, \und a\, j\, \und b}\,(q, \tilde q)
= \int du\, dv \, \eps_{\und b\, \und a}\, u^1_{(i} u^{-1}_{j)}\,
\frac{\partial^2 {\cal L}^{1,0\; \und d}}
{\partial q^{1,0\; \und d}\, \partial q^{0,1\; \und k}}\,  v^1_c\,.
\label{DirII}
\ee
Integrating by parts with respect to the harmonic derivative $\partial^{0,2}$
($v^1_a = \partial^{0,2} v^{-1}_a$) and using the constraint
\be
\frac{\partial {\cal L}_0^{2,2}}{\partial \qa}
= \partial^{0,2} {\cal L}^{1,0}_{\,\und a}\,, \nn
\ee
we immediately find that \p{DirII} is vanishing.

Once again, repeating the same analysis for the fermionic terms, one can be
convinced that the similar splitting into a sum of independent actions
takes place for these terms as well. Hence, this phenomenon persists at
the full superfield level for the multiplets $\hat q^{\,\und i\, a}$
and $\hat q^{\,i\, \und a}$\,, like in the previously considered case
of the multiplets $\hat q^{\,i\, a}$ and $\hat q^{\,i\, \und a}$\,.

\newpage

\noindent{\it Adding more multiplets.}\\

As the last topic of this Subsection, let us briefly discuss the case
when the Lagrangian ${\cal L}^{2,2}$ is originally allowed to depend on three different types
of the twisted multiplet, say, on the following triple of superfields
\be
(\,q^{1,1}\,,\quad \hat q^{\,1,0\;\und a}\,,\quad \hat q^{\,0,1\; \und i}\,)\,.
\label{triple}
\ee

Firstly we note that the corresponding Lagrangian should have the same
structure as in (\ref{SD}), because the inclusion of the analytic superfield
$q^{1,1}$ as an additional functional parameter in the superpotentials in \p{SD}
is harmless for its form.
The requirement that the action for the triple \p{triple} is invariant under
the supersymmetry transformations leads to the system of constraints which
looks the same as in the case of pair of the multiplets $\hat q^{\,1,0\;\und a}$
and $\hat q^{\,0,1\; \und i}$\,, i.e. are given by eqs. (\ref{cdoubl}).
The straightforward calculation of the component action with making use of
the constraints (\ref{cdoubl}) leads to the following conclusions about its structure: \\

\noindent {\it{(i).}} The auxiliary boson part of the action is reduced to
a sum of three pieces, and each of these pieces corresponds to the auxiliary
boson part of the relevant separate multiplet.\\

\noindent {\it{(ii).}} As in the previous cases, the physical boson part of the
action contains some off-diagonal terms with $x$-derivatives. \\

An inspection of these mixed terms shows that their structure is similar
to that we met in the two previous cases. Although there appear some
extra pieces arising e.g. from the action  of the pair of
$(q^{1,1}, \hat q^{\,0,1\; \und i})$,
the physical boson part can be fully diagonalized as before, by using
the constraints (\ref{cdoubl}). The result of this procedure can be
schematically written as the splitting of the action into a sum of
the three pieces for the separate multiplets
\be
S(q^{1,1}\,,\; \hat q^{\,1,0\;\und a}\,,\; \hat q^{\,0,1\; \und i})
= S(q^{1,1}) + S(\hat q^{\,1,0\;\und a}) + S(\hat q^{\,0,1\; \und i})
\ee
where the metric and torsion terms can still depend on all three sets of
physical bosons. However, using the constraints once again, it is easy to
demonstrate that both scalar functions and torsion potentials in every piece
can bear dependence only on the physical bosons of its ``own'' multiplet.
The proof of separation of the fermionic terms follows the same routine.

In a similar way one can prove the separation property for any number of
non-equivalent multiplets.

\subsection{Potential terms for $\hat q^{\,1,0\; \und a}$ and
$\hat q^{\,0,1\; \und i}$}

In Sect. 5 we constructed the potential terms for separate twisted multiplets.
Here we show the existence of mixed mass terms which involve different types
of multiplets. These terms are in fact of the same form as those given in
\cite{GI}, \cite{G1} and can be constructed only for multiplets belonging to
the same ``self-dual'' pair. A new finding is the general form of the relevant
scalar potential arising after elimination of the auxiliary fields.
This potential and the accompanying Yukawa-type fermionic terms are the only
mixed interaction of twisted multiplets of different types compatible
with ${\cal N}=(4,4)$ supersymmetry.

The candidate mixed mass--terms can be written in the analytic
superspace in the following form
\be
S^M = M \int \mu^{-2,-2}\, C^{\,0,0}_{\und i\; \und k\; \und a\; \und b}\,
\theta^{1,0\; \und k}\, \theta^{0,1\; \und b}\, \qa\, \qb \,,
\label{MD}
\ee
\be
S^M_1 = M_1 \int \mu^{-2,-2}\, C^{\,0,0}_{\und i\; \und k\; \und a\; \und b}\,
\theta^{1,0\; \und k}\, \theta^{0,1\; \und b}\, q^{1,1}\, \tilde q^{\, \und i\, \und a}\,,
\label{MD1}
\ee
\be
S^M_2 = M_2 \int \mu^{-2,-2}\, C^{\,-1,0}_{\und k\; \und a\; \und b}\,
\theta^{1,0\; \und k}\, \theta^{0,1\; \und b}\, q^{1,1}\, \qa\,,
\label{MD2}
\ee
etc. All terms of this kind with higher powers of the involved
superfields can be shown to vanish because of the corresponding harmonic
constraints. As for the terms \p{MD} - \p{MD2}, only those given in
(\ref{MD}) and (\ref{MD1}) can respect ${\cal N}=(4,4)$ supersymmetry for
non-vanishing harmonic constants $C$. The term \p{MD2} and any other
similar term involving superfields from different ``self-dual'' pairs
(e.g. from $\hat q^{\,1,0\;\und a}$ and $\hat q^{\,\und i\,\und a}$) can easily
be shown to vanish as a consequence of the requirement of supersymmetry.

Without loss of generality, let us restrict our consideration to
the mass term (\ref{MD}).
Computing the supersymmetry variation of the action (\ref{MD}),
it is easy to find that \p{MD} is invariant provided the harmonic
constants $C$ satisfy the following conditions
\be
\partial^{2,0} C^{\,0,0}_{\und i\; \und a\; \und k\; \und b} =0\,, \quad
\partial^{0,2} C^{\,0,0}_{\und i\; \und a\; \und k\; \und b} =0\,, \quad
C^{\,0,0}_{\und i\; \und a\; \und k\; \und b} =
\eps_{\und i\, \und k}\, \eps_{\und a\, \und b}\,.
\ee
After performing the integration over Grassmann and harmonic variables
in (\ref{MD}), one finds the {\it{off-shell}} component form of this term:
\be
S^M = -\frac{M}{4} \int d^2x\; \{\, q^{\und k\, b}\; F_{\und k\, b}
+ q^{i\, \und a}\; F_{i\, \und a}\, \}\,.
\label{mdoubl}
\ee

For sake of simplicity, we shall consider only bosonic part of the full component
on-shell action. After eliminating the auxiliary fields $F_{\und k\, b}$ and
$F_{i\, \und a}$ in the sum $S_{(a)}^{bos} + S_{(b)}^{bos} + S^M$
where $S_{(a)}^{bos}$ and $S_{(b)}^{bos}$ are bosonic parts of the component
sigma model actions for $\hat q^{\,1,0\; \und a}$ and $\hat q^{\,0,1\; \und i}$
($S_{(a)}^{bos}$ is given by eq. (\ref{comp.aux22}) and $S_{(b)}^{bos}$ has
a similar form), the induced {\it{on-shell}} scalar potential term reads
\be
S^{pot} =  \frac{M^2}{8} \int d^2x\; \{\,G^{-1}\, q^{\und k\, b}\, q_{\und k\, b}
+ \tilde G^{-1}\, q^{i\, \und a}\, q_{i\, \und a}\, \}.
\label{POTm}
\ee
Here $G = G(q^{i\, \und a})$ and $\tilde G = \tilde G(q^{\und k\, b})$
are the bosonic scalar metrics of the $\hat q^{\,1,0\; \und a}$ and
$\hat q^{\,0,1\; \und i}$ multiplets (they are defined by eqs. \p{Gq2},
\p{gq2} and by similar ones for $q^{\und k\, b}$)\,.
Thus we see that in the general interaction case
corresponding to nontrivial metric functions, the potential \p{POTm}
contain mixed couplings of two different twisted multiplets. It is easy to
restore the fermionic terms as well. We see that \p{POTm} yields mass terms
for the involved fields even in the case of constant functions $G$
and $\tilde G$, i.e. if one starts from the free kinetic actions of
the twisted multiplets considered. This is a difference from similar
superfield terms \p{Smass2}, \p{Smass1} for single multiplets.

The most general {\it{off-shell}} mass term for the pair of multiplets
$(\hat q^{\,1,0\;\und a}\,, \,\hat q^{\,0,1\; \und i})$ can be written
as a sum of the following three pieces
\be
S^M_{(1+2)} = S^M + S^m_{(a)} + S^m_{(b)}
\label{MASS}
\ee
where
\be
S^m_{(b)} = - \frac{m_{(b)}}{4} \int d^2x \; C^{k\, \und b}\, F_{k\, \und b}\,
\label{mass1.off}
\ee
and two other terms in (\ref{MASS}) are given by the expressions (\ref{mdoubl})
and (\ref{mass.off}). After eliminating the auxiliary fields $F_{\und i\, a}$
and $F_{k\, \und b}$ in the sum $S_{(a)}^{bos} + S_{(b)}^{bos} + S^M_{(1+2)}$\,,
the most general {\it{on-shell}} potential part of the action is obtained
in the form
\bea
S^{Pot} &=& \frac{1}{8} \int d^2x \; \{\, G^{-1}\, ( m^2_{(a)}\, C^{\und k\, b}\, C_{\und k\, b}
+ 2\, m_{(a)}\, M\, C^{\und k\, b}\, q_{\und k\, b}
+ M^2 q^{\und k\, b}\, q_{\und k\, b}) \nn \\ [0.3cm]
&+& \tilde G^{-1}\,
( m^2_{(b)}\, C^{i\, \und a}\, C_{i\, \und a}
+ 2\, m_{(b)}\, M\,  C^{i\, \und a}\, q_{i\, \und a}
+ M^2 q^{i\, \und a}\, q_{i\, \und a}) \,\}\,.
\eea

\section{Conclusions}

In this paper we extended our previous analysis of the manifestly ${\cal N}=(4,4)$
supersymmetric off-shell description of the twisted multiplet $q^{1,1}$
in the $SU(2)\times SU(2)$ HSS \cite{IS} to  the case of other three types of
such a multiplet, which differ in assignments of their component fields with
respect to the full R-symmetry group $SO(4)_L\times SO(4)_R$ of ${\cal N}=(4,4)$
$2D$ Poincar\'e superalgebra.
We constructed off-shell superfield actions for each of these new multiplets
in the analytic subspace of the $SU(2)\times SU(2)$ HSS and, as an example,
discussed the special case of superconformally invariant ${\cal N}=(4,4)$ superextension
of the $SU(2) \times U(1)$ group manifold WZNW sigma model associated with
one of these multiplets (represented by  the analytic superfunction $\qa$).
Since the Lagrangians of these alternative twisted multiplets are expressed in
terms of the harmonic analytic superfunctions having non-standard transformation
properties under ${\cal N}=(4,4)$ supersymmetry, the requirement that the corresponding
actions are supersymmetric leads to certain constraints on the structure of
the Lagrangians. Using these constraints, we were able to show that the bosonic
target geometries of sigma models for the new multiplets are of the same sort as
in the case of the $q^{1,1}$ multiplet considered in \cite{IS}. We also discussed
massive extensions of general sigma model actions of the multiplets
$\hat q^{\,1,0\; \und a}$ and $\hat q^{\,\und i\, \und a}$ and, as an example,
presented a massive deformation of the conformal WZNW action of the
$\hat q^{\,1,0\; \und a}$ multiplet. Like the sigma model actions, the mass terms
for the new multiplets reveal the same structure as those for the $q^{1,1}$ multiplet.

The basic new findings of our study are related to the analysis of the options
when two or more multiplets of different sort are allowed to interact with each
other via the sigma model- or/and mass term-type analytic Lagrangians.
We have found that ${\cal N}=(4,4)$ supersymmetry requires the general sigma model action
of any pair of such multipelts to reduce to a sum of sigma model actions of
separate multiplets, and this phenomenon persists in the cases when a larger number
of different multiplets is involved into the game. The only possibility to arrange
mutual interactions of the twisted multiplets of different types is via the appropriate
mixed mass terms. The latter are bilinear in the multuplets belonging to the same
``self-dual'' pair which is characterized by the property that the $SU(2)$ assignments
of the physical and auxiliary bosonic fields of the involved multiplets are
complementary to each other. The multiplets belonging to different such pairs, can
interact with each other neither via sigma model type actions nor via mass terms.
For a ``self-dual'' pair of twisted multiplet we have given the most general form
of the scalar bosonic potential which arises as a result of eliminating auxiliary
fields in the sum of general sigma model actions of these multiplets and three
possible mass terms, including the mixed one.

One of the possible directions of extending the study undertaken in this paper is
to couple the considered models to conformal ${\cal N}=(4,4)$ supergravity in
the $SU(2)\times SU(2)$ HSS formulation of ref. \cite{bi1}. On this way one can
hope to discover new off-shell versions of Poincar\'e ${\cal N}=(4,4)$ supergravity,
with the new types of twisted multiplet as superconformal compensators.
One more interesting task is to study a possible effect of incorporating
the additional twisted multiplets into more general HSS sigma models with
non-commuting left and right quaternionic structures on the target space \cite{EI}.
More technical work which is now under way \cite{IS4} is to repeat the analysis of
the present paper in terms of ${\cal N}=(2,2)$ superfields. The ${\cal N}=(2,2)$ superspace
language is used in many studies of ${\cal N}=(2,2)$ and ${\cal N}=(4,4)$ supersymetric sigma
models with torsion and it is capable to make some proofs and observations of
the present paper more tractable and clear.

\section*{Acknowledgements}
We acknowledge a partial support from INTAS grant, project No 00-00254,
and RFBR grant, project  No 03-02-17440. The work of E.I. was
also supported by the RFBR-DFG grant No 02-02-04002, and a grant of the
Heisenberg-Landau program. E.I. thanks S. Bellucci for interest in this study
at its early stages. A.S. is thankful to S. Krivonos and A. Pashnev for helpful
comments and discussions.

\newpage

\section*{Appendix} \setcounter{equation}{0}
\renewcommand{\theequation}{A.\arabic{equation}}

\noindent{\it Analytic superspace integration measure}

\be
\int d^2 \theta^{1,0}\, d^2 \theta^{0,1} =
\frac{1}{16}\, \eps^{\und k\, \und n}\, \eps^{\und c\, \und d}\,
\frac{\partial}{\partial \theta^{1,0\, \und k}}\, \frac{\partial}{\partial \theta^{1,0\, \und n}}\,
\frac{\partial}{\partial \theta^{0,1\, \und c}}\, \frac{\partial}{\partial \theta^{0,1\, \und d}}\,,
\ee
\be
\int d^2 \theta^{1,0}\, d^2 \theta^{0,1}\,
(\theta^{1,0})^2 (\theta^{0,1})^2 = 1\,.
\ee

\vspace{0.4cm}
\noindent{\it Complete solution to the constraints (\ref{harm.con2}) -- (\ref{harm.con4})}

\bea
g^{1,1} &=& - \beta^{i\,a} u^1_i v^1_a + \ti F^a_{\und i} v^1_a +
2i \ta \partial_{--} q^i_{\und a} u^1_i \nn \\
&-& 2i \ti \ta \partial_{--}\alpha_{\und i\, \und a}
+ i (\theta^{1,0})^2 \partial_{++} \beta^{i\,a} u^{-1}_i v^1_a
+ i (\theta^{0,1})^2 \partial_{--}\beta^{i\,a} u^1_i v^{-1}_a  \nn\\
&-& i \ti (\theta^{0,1})^2 \partial_{--} F^a_{\und i} v^{-1}_a
+ 2 (\theta^{1,0})^2 \ta \partial_{++} \partial_{--} q^i_{\und a} u^{-1}_i \nn\\
&+& (\theta^{1,0})^2 (\theta^{0,1})^2 \partial_{++} \partial_{--}
\beta^{i\,a} u^{-1}_i v^{-1}_a\,,
\eea
\bea
f^{1,1} &=& - \rho^{i\,a} u^1_i v^1_a
+ 2i \ti \partial_{++} q^a_{\und i} v^1_a  - \ta F^i_{\und a} u^1_i  \nn\\
&+& 2i \ti \ta \partial_{++}\gamma_{\und i\, \und a}
+ i (\theta^{1,0})^2 \partial_{++} \rho^{i\,a} u^{-1}_i v^1_a
+ i (\theta^{0,1})^2 \partial_{--}\rho^{i\,a} u^1_i v^{-1}_a  \nn\\
&+& 2 \ti (\theta^{0,1})^2 \partial_{++} \partial_{--} q^a_{\und i} v^{-1}_a
+ i (\theta^{1,0})^2 \ta \partial_{++} F^i_{\und a} u^{-1}_i \nn\\
&+& (\theta^{1,0})^2 (\theta^{0,1})^2 \partial_{++} \partial_{--}
\rho^{i\,a} u^{-1}_i v^{-1}_a\,,
\eea
\bea
h^{0,1 \und i} &=& \xi^{\und i\, a} v^1_a + \ti F^{i\,a} u^{-1}_i
v^1_a + 2i \ta \partial_{--} q_{\und a}^{\und i}
- i (\theta^{0,1})^2 \partial_{--} \xi^{\und i\, a} v^{-1}_a  \nn \\
&+& 2i \ti \ta \partial_{--} \psi^i_{\und a} u^{-1}_i - i \ti
(\theta^{0,1})^2 \partial_{--} F^{i\,a} u^{-1}_i v^{-1}_a \,,
\eea

\bea f^{1,0 \und a} &=& \psi^{i\, \und a} u^1_i + 2i \ti
\partial_{++} q^{\und a}_{\und i}
- \ta F^{i\,a} u^1_i v^{-1}_a - i (\theta^{1,0})^2 \partial_{++} \psi^{i\, \und a} u^{-1}_i \nn\\
&-& 2i \ti \ta \partial_{++} \xi^a_{\und i} v^{-1}_a + i
(\theta^{1,0})^2 \ta \partial_{++} F^{i\,a} u^{-1}_i v^{-1}_a \,,
\eea
\bea
t^{1,1} &=& F^{i\,a} u^1_i v^1_a - 2i \ti \partial_{++} \xi_{\und i}^a v^1_a
+ 2i \ta \partial_{--} \psi^i_{\und a} u^1_i \nn\\
&-& i (\theta^{1,0})^2 \partial_{++} F^{i\,a} u^{-1}_i v^1_a - i
(\theta^{0,1})^2 \partial_{--} F^{i\,a} u^1_i v^{-1}_a
+ 4 \ti \ta \partial_{++} \partial_{--} q_{\und i\, \und a} \nn \\
&-& 2 \ti (\theta^{0,1})^2 \partial_{++} \partial_{--} \xi_{\und i}^a v^{-1}_a
+ 2 (\theta^{1,0})^2 \ta \partial_{++} \partial_{--} \psi^i_{\und a} u^{-1}_i  \nn \\
&-& (\theta^{1,0})^2 (\theta^{0,1})^2 \partial_{++} \partial_{--}
F^{i\,a} u^{-1}_i v^{-1}_a\,.
\eea

\end{document}